\documentclass[twocolumn,prb]{revtex4}
\usepackage{graphicx}
\usepackage{amssymb}

\newcommand{\be}{\begin{equation}}
\newcommand{\ee}{\end{equation}}
\newcommand{\br}{{\bf r}}
\newcommand{\brp}{{\bf r}^{\prime}}

\newcommand{\bx}{{\bf x}}

\newcommand{\prt}{\partial}

\newcommand{\hH}{\hat{H}}

\newcommand{\hO}{\hat{O}}

\newcommand{\hpsi}{\hat{\psi}}
\newcommand{\psia}{\hat{\psi}}
\newcommand{\psic}{\hat{\psi}^{\dagger}}
\newcommand{\hpsid}{\hat{\psi}^{\dagger}}

\newcommand{\half}{\frac{1}{2}}

\newcommand{\hU}{\hat{U}}
\newcommand{\Tr}{\textrm{Tr}}
\newcommand{\tr}{\textrm{tr}}
\newcommand{\la}{\langle}
\newcommand{\ra}{\rangle}
\newcommand{\bea}{\begin{eqnarray}}
\newcommand{\eea}{\end{eqnarray}}

\begin{document}

\title{Total energies from variational functionals of the Green function
and the renormalized four-point vertex}
\author{Robert van Leeuwen}
\author{Nils Erik Dahlen}
\author{Adrian Stan}
\affiliation{Rijkuniversiteit Groningen, Theoretical Chemistry,
Materials Science Centre, 9747AG, Nijenborgh 4, Groningen, The
Netherlands}

\date{\today}

\begin{abstract}
We derive variational expressions for the grand potential or action in terms of the
many-body Green function $G$ which describes the propagation
of particles and the renormalized four-point vertex $\Gamma$ which describes the
scattering of two particles in many-body systems. 
The main ingredient of the variational functionals is a term we denote
as the $\Xi$-functional which plays a role analogously to the
usual $\Phi$-functional studied by Baym
(G.Baym, Phys.Rev. {\bf 127}, 1391 (1962)) in connection with the conservation laws in
many-body systems. We show that any $\Xi$-derivable theory is also $\Phi$-derivable
and therefore respects the conservation laws.
We further set up a computational scheme to obtain 
accurate total energies from our variational functionals without having to solve
computationally expensive sets of self-consistent equations. The input of the
functional is an approximate Green function $\tilde{G}$ and 
an approximate four-point vertex 
$\tilde{\Gamma}$ obtained
at a relatively low computational cost. The variational property of the functional
guarantees that the error in the total energy is only of second order 
in deviations of the input Green function and vertex from the self-consistent ones
that make the functional stationary.
The functionals that we will consider for practical applications correspond to infinite
order summations  of ladder and exchange diagrams and are therefore particularly
suited for applications to highly correlated systems. Their practical evaluation is 
discussed in detail.
\end{abstract}

\maketitle

\section{Introduction}
Total energy calculations play an important role in condensed matter
physics and quantum chemistry.
For solid state physicists they are essential in predicting 
structural changes and bulk moduli in
solids. In chemistry molecular bonding curves and potential 
energy surfaces are essential to understand phenomena like
molecular dissociation and chemical reactions.
However, accurate total energy calculations are
notoriously difficult and computationally demanding. In quantum chemistry there are
advanced wavefunction methods like configuration interaction and
coupled cluster theory~\cite{Helgaker:book} to calculate energies
but they can only be applied to relatively small
molecules. In solid state physics most total energy calculations
for crystals or surfaces 
are based on density functional theory~\cite{DreizlerGross:book} where
the density functionals are mostly based on the local density approximation (LDA) and
generalized gradient approximations (GGA)~\cite{KurthPerdew:2003}. These functionals 
have had great success
but there are many cases where the functionals fail, in which case there
is no clear systematic route to improvement.
We have therefore recently advanced a different scheme
which involves variational energy functionals of the many-body Green
function and applied it succesfully to calculate total energies of atoms, 
molecules~\cite{DahlenvonBarth:JCP04,DahlenvonBarth:PRB04,Dahlenetal:IJQC05,DahlenvanLeeuwen:JCP05,Dahlenetal:PRA06}
and the electron gas~\cite{ABL:IJMP99}. 
A variety of such functionals can be systematically constructed
using diagrammatic perturbation theory in which the
different functionals correspond to different levels of perturbation theory.
For these functionals
we use input Green functions that
are relatively easy to obtain at low computational cost, 
for instance from a local density or Hartree-Fock calculation.
The variational property of the functional then assures that the errors in the
energy are only of second order in the difference
between our approximate Green function and the actual Green
function that makes the functional stationary.
This is the essential feature that allows one to obtain accurate total energies
at a relatively low computational cost~\cite{Dahlenetal:PRA06}.
The remaining question is then how to select
approximate variational functionals that yield good total energies.

In a diagrammatic expansion in many-body
perturbation theory the building blocks
are the Green function line $G$, which describes the
propagation of particles and holes, and the interaction line $v$
which in electronic systems is represented by the
Coulomb repulsion between the electrons.
From this diagrammatic structure one can proceed
to construct variational functionals in several ways.
First of all we can renormalize the Green function 
lines. This leads to a functional that has been
studied by Luttinger and Ward~\cite{LuttingerWard:PR60} and
leads to a functional we will call the $\Phi$-functional
$\Phi [G,v]$, depending on the dressed Green function and the
bare two-particle interaction $v$. The Luttinger-Ward functional has been
applied, with great success, to the calculation of total energies
of the electron gas~\cite{Hindgren:thesis,ABL:IJMP99}, and atoms and 
molecules
~\cite{DahlenvonBarth:JCP04,DahlenvonBarth:PRB04,Dahlenetal:IJQC05,DahlenvanLeeuwen:JCP05,Dahlenetal:PRA06,Aryasetiawan:PRL02}.
This type of functionals has also received considerable attention
for Hubbard lattice type systems~\cite{Janis:condmat,Janis:PRB99,Aryasetiawan:PRL02,Potthoff:EPJB03,Potthoff:PRL03,Janis:JPC03}.
Apart from renormalization of the Green function lines, 
we can also decide to renormalize the
interaction lines by replacing the bare interaction
by a dynamically screened one, usually denoted by $W$. This leads to
the functional $\Psi [G,W]$ that was investigated in a
paper by Hedin~\cite{Hedin:PR65} and later more completely 
by Almbladh et al.\cite{Hindgren:thesis,ABL:IJMP99}
and which has been applied with succes to calculations of the
total energy of the electron 
gas~\cite{Hindgren:thesis,ABL:IJMP99} and atoms~\cite{DahlenvonBarth:PRB04}.
This type of functionals has also received considerable attention
in the Dynamical Mean Field Theory (DMFT) 
community~\cite{ChitraKotliar:PRB01,SavrasovKotliar:PRB04,Tong:PRB05,Kotliaretal:RMP}.
The natural place to use this functional
is in extended systems in which screening of the long range
Coulomb interaction is essential.
Finally there is also the possibility to renormalize
the four-point vertices and replace them by a
renormalized four-point vertex $\Gamma$. 
In this work we will concentrate on this type of functionals.

The natural place for variational functionals of the Green function $G$ and 
the four-point vertex $\Gamma$ is in systems
where short range correlations play an important role such as in
highly correlated systems. Such a type of theory was recently
discussed in work of Jani{\v s}~\cite{Janis:condmat,Janis:PRB99} on the
Hubbard model in which it was demonstrated how to derive
the so-called parquet approximation from a functional of the
Green function and the four-point vertex.  
Furthermore Katsnelson and Lichtenstein~\cite{KatsnelsonLichtenstein:EPJB02} 
have considered the electronic structure of correlated metals in which 
the building blocks of the theory are an approximate $T$-matrix
and a bare or noninteracting Green function 
(or a bare Green function in an effective correlated medium 
when using Dynamical Mean Field Theory\cite{Georges:RMP96}).
For describing the structural properties of such materials
it would therefore be of great importance to be able
to calculate the total energy from 
variational energy expressions in terms of the Green function
and the four-point vertex where we use an approximate $G$ and $\Gamma$
as an input. The variational property then guarantees that
the errors in the energy are only of second order in the deviations
of the input Green function and vertex from the true quantities that
make the functional stationary.

The construction of energy functionals 
in terms of $G$ and $\Gamma$ is most naturally done by the
use of the Hugenholtz diagram 
technique~\cite{Hugenholtz:57,Nozieres:book, NegeleOrland:book,BlaizotRipka:book}
which has the bare four-point vertex as a diagrammatic building block. This 
procedure has been carried out in the early 1960s by 
De Dominicis~\cite{DeDominicis:JMP62,DeDominicis:JMP63} and later 
in more generality by De Dominicis and 
Martin~\cite{DeDominicisMartin:JMP(1)64,DeDominicisMartin:JMP(2)64} and
leads to a functional we will call the $\Xi$-functional
$\Xi [G, \Gamma]$.  In the latter works the
derivation has been carried out for a very general many-body system with
not only one- and two-body interactions but also with
$\frac{1}{2}$-body and $\frac{3}{2}$-body interactions
that describe Bose-condensed and superconducting phases.
Unfortunately this leads to rather involved equations
and disguises the simpler case in which there are
only one and two-body interactions. For instance, in the
general Bose-condensed and super-conducting case no
particle-particle and particle-hole contributions to
the four-point vertex can be distinguished. The work of De Dominicis and
Martin was aimed at demonstrating that one could express all 
thermodynamic quantities completely in terms of distribution functions
rather than at a practical application of the formalism. 
In their work there is, therefore, no discussion of approximate
functionals and of ways of evaluating them.
However, nowadays the functionals can be subjected to numerical
computation and it is therefore timely to discuss the formalism
from this point of view and to present computational schemes
to evaluate the functionals to calculate total energies. 
This is exactly the purpose of this work.

If we consider the first of the two papers of De Dominicis and 
Martin~\cite{DeDominicisMartin:JMP(1)64} we see that they
use a purely algebraic approach to construct their functional which is not
capable of displaying its full structure. Their
 second paper~\cite{DeDominicisMartin:JMP(2)64} uses
a purely diagrammatic approach to derive in much more detail
the structure of the functional but the
derivation is quite difficult due to numerous intricate topological
theorems that need to be discussed in order to avoid double
counting of the diagrams.
However, we found that a combination of both methods discussed
in these two papers leads to a much quicker derivation of the
final results.
Therefore, in this work we derive, in a as simple as possible manner,
a variational energy or action functional for normal systems using a 
purely algebraic method in combination with
a diagrammatic analysis.
We use, however, one generalization of the formalism of
DeDominicis and Martin: 
since the Green functions are 
generated by differentiation of our functionals with
respect to time-nonlocal potentials, 
the most natural framework to use is the
Keldysh Green function 
technique~\cite{Keldysh:JETP65,Danielewicz:AP84,Bonitz:book,condmat}.  
We therefore consider generally
time-dependent systems that
are initially in thermodynamic equilibrium. 
This has two other advantages. Firstly it allows for
an elegant discussion of conservation laws which,
as was shown by Baym~\cite{Baym:PR62}, are closely connected to
$\Phi$-derivability. Such conservation laws were earlier
discussed 
for variational energy and action functionals within the $\Phi$- and $\Psi$-formalism 
 in connection
with time-dependent density-functional theory~\cite{Ulfetal:PRB05}.
In particular we will in this paper show that also $\Xi$-derivable theories are conserving.
Secondly, the use of finite temperature allows for an
elegant treatment of the boundary conditions on
the Green functions. These are, for instance, essential
in going from the equations of motion for the Green
function to the Dyson equation which will play an important
role in our derivations.

The paper is divided as follows.
We first discuss some definitions that form the basis
of our subsequent analysis. We then derive self-consistent
equations that relate the Green function and the renormalized
four-point vertex. Then we provide a general construction of the
variational functional using purely algebraic methods and
we subsequently analyze the structure of the functional
using diagrammatic methods. We then briefly discuss the
conserving properties of the functional. Finally we discuss approximate
functionals with details for their practical evaluation
and present our conclusions and outlook.

\section{Defining equations}

In the following we will consider a many-body system initially in thermodynamic
equilibrium. At an initial time $t_0$ the system is subjected to
a time-dependent external field.
The Hamiltonian of the system in an time-dependent external
potential $w(\bx t)$ is (in atomic units) given by
\be
\hH (t) = \hat{h} (t) + \hat{V}
\ee
where in the usual second quantization notation
the one- and two-body parts of the Hamiltonian are given by
\begin{eqnarray}
\hat{h} (t) &=&   \int d\bx \hpsid (\bx)
h (\bx t) \hpsi (\bx)  
\label{eq:ho}\\
\hat{V} &=& \half \int d\bx d\bx' v(\br, \brp) \hpsid (\bx)
\hpsid (\bx') \hpsi (\bx') \hpsi (\bx) .
\end{eqnarray}
Here $\bx= (\br, \sigma)$ is a space-spin coordinate.
The two-body interaction will usually be taken to be
a Coulombic repulsion, i.e. $v(\br,\br')=1/|\br -\br'|$.
The one-body part of the Hamiltonian 
has the explicit form
\be
h (\bx t) =  -\half  \nabla^2  + w(\bx t) - \mu .
\label{eq:ho2}
\ee
We further introduced the chemical potential $\mu$
in the one-body part of the Hamiltonian of Eq.(\ref{eq:ho2})
in anticipation of a finite temperature treatment of the system.
We first consider the expectation value of an
operator $\hat O$ for the case that the system is initially in an 
equilibrium state before a certain time $t_0$. 
For $t < t_0$ the expectation value of operator $\hat O$ in the
Schr\"odinger picture is then
given by $\langle \hat O \rangle = \Tr \{ \hat \rho \hat O \}$
where $\hat \rho = e^{-\beta \hH_0} / \Tr { e^{-\beta \hH_0}}$ is the
density matrix and $\hH_0$ is the time-independent Hamiltonian that
describes the system before the perturbation is switched on.
We further defined $\beta=1/k_B T$, with $k_B$ the Boltzmann constant,
to be the inverse 
temperature, and the trace involves a summation
over a complete set of states in the Hilbert space.
After we switch on the field the expectation value becomes
a time-dependent quantity given by
\be
\langle \hat O \rangle (t) = \Tr \left\{ \hat \rho \hat O_H (t) \right\}
\ee
where 
$\hat O_H(t) = \hU(t_0, t) \hat O(t) \hU(t, t_0)$ is
the operator in the Heisenberg picture.
The evolution operator $\hat U$ of the system is
defined as the solution to the equations
\bea
i\partial_t \hU(t, t') &=& \hH(t) \hU(t,t') 
\label{eq:umot1} \\
 i\partial_{t'} \hU(t,t') &=& -\hU(t,t')\hH(t')
 \label{eq:umot2}
\eea
with the boundary condition $\hU(t,t)=1$ .
The formal solution of Eq.~(\ref{eq:umot1}) can be obtained by integration to yield
the time-ordered expression $\hU(t,t') = T \exp{(-i\int_{t'}^t d\tau \hH (\tau) )}$ for $t>t'$
with a similar expression with anti-chronological time-ordering
for $t' >t$.
The operator $e^{-\beta \hH_0}$
can now be regarded as an evolution operator in imaginary time, i.e.
$\hU(t_0-i\beta,t_0)=e^{-\beta\hH_0}$, 
if we define $\hH(t)$ to be equal to $\hH_0$ on the contour running straight from
$t_0$ to $t_0-i\beta$ in the complex time plane. We can therefore rewrite
our expression for the expectation value as
\be
\langle \hat O \rangle (t) = \frac{ \Tr \left\{ \hU (t_0-i\beta, t_0) \hU (t_0 , t) \hat O
\hU (t, t_0) \right\} }{ \Tr \left\{ \hU (t_0-i\beta, t_0) \right\} }
\label{eq:average}
\ee
If we read the time arguments of the evolution operators in the numerator
of this expression from left to right we may say that the system
evolves from $t_0$ along the real time axis to $t$ after which the
operator $\hat O$ acts. Then the system evolves back along the real axis
from time $t$ to $t_0$ and finally parallel to the imaginary axis from
$t_0$ to $t_0-i \beta$. 
\begin{figure}
\begin{center}
\includegraphics[width=7.0cm]{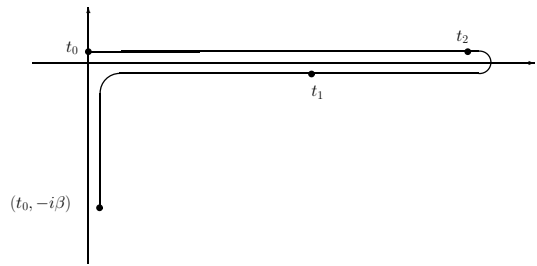}
\end{center}
\caption{The Keldysh contour drawn in the complex time plane}
\label{fig:contour}
\end{figure}
This observation motivates us to define
the following action functional
(compare with the action functionals used in
Refs.\cite{Baym:PR62,vanLeeuwen:PRL98})
\be
Y =  i \ln \Tr \left\{ \hU (t_0 -i\beta, t_0) \right\} , 
\label{eq:action}
\ee
where 
we define the evolution operator on the contour as
\be
\hU (t_0-i\beta, t_0) = T_C \exp (-i \int dt\hH (t) ).
\label{eq:contevol}
\ee
Here the integral is taken on the contour and $T_C$
denotes time-ordering along the contour~\cite{Danielewicz:AP84,condmat}.  
When we evaluate this quantity for the equilibrium system we
see that
\be
iY = - \ln \Tr \left\{ e^{-\beta \hH_0} \right\} = \beta \Omega
\ee
where $\Omega$ is the grand potential. Therefore the total energy $E$
of the system is obtained from the zero-temperature limit
\be
\lim_{T \rightarrow 0} \frac{iY}{\beta} = \lim_{T \rightarrow 0} \Omega = E- \mu N
\ee
where $N$ denotes the number of particles in the system.
Let us now see how this
functional can be used as a generating functional by making variations
with respect to parameters in the Hamiltonian. To do this one needs to consider
changes in the evolution operator $\hat U$ which are readily evaluated using Eqs.(\ref{eq:umot1})
and (\ref{eq:umot2}).
For instance, when we make a perturbation $\delta \hat V (t)$ in the Hamiltonian
we have using Eq.(\ref{eq:umot1})
\be
i \partial_t \, \delta \hat U (t,t') = \delta \hat V (t) \hat U (t,t') + \hat H (t) \delta U (t,t')
\ee
with a similar differential equation with respect to $t'$ and boundary condition
$\delta \hat U (t,t)=0$. The solution to this equation is given by
\be
\delta \hat U(t,t') = 
-i\int_{t'}^t d\tau \hat U(t, \tau) \delta \hat V(\tau) U(\tau, t')
\label{eq:solution}
\ee 
from which variations in the action can be calculated.
For instance, if we choose the perturbation to be
a time-dependent and spatially nonlocal potential of the form
\be
 \delta \hat V (t) =\int d\bx_1 d\bx_2 \, \delta u (\bx_1, \bx_2, t) \psic (\bx_1)
\psia (\bx_2)
\ee
we obtain the time-dependent one-particle density matrix as
a functional derivative with respect to $Y$
\be
\la \psic (\bx_1) \psia (\bx_2) \ra (t) =  \frac{\delta Y}{ \delta u(\bx_1,\bx_2, t)} .
\label{eq:1mat}
\ee
Similarly, when we consider a time-dependent two-body potential of
the form
\bea
\delta \hat V (t) = \int d(\bx_1 \bx_2 \bx_3 \bx_4)
\ \delta V(\bx_1 \bx_2 \bx_3 \bx_4,t) \nonumber \\
 \times 
\psic (\bx_1) \psic (\bx_2) \psia (\bx_3) \psia (\bx_4)
\eea
we obtain the time-dependent two-particle density matrix as a derivative
\bea
\langle \psic (\bx_1) \psic (\bx_2) \psia (\bx_3) \psia (\bx_4) \rangle (t) \nonumber \\
= \frac{\delta Y}{\delta V(\bx_1 \bx_2 \bx_3 \bx_4,t)}
\label{eq:2mat}
\eea
Note that in order to derive Eqs.(\ref{eq:1mat}) and (\ref{eq:2mat}) 
we had to make variations $\delta u$
and $\delta V$ for time-variables on the contour. After the variation is made 
all observables are, of course, evaluated for physical quantities that are the same
on the upper and lower branch of the contour.
In the remainder of the paper we will heavily use the action functional
as a generating functional for the many-body Green functions.
To do this we have to generalize the time-local potentials $u$ and $V$ to
time-nonlocal ones, such that the derivatives of $Y$ with respect to these potentials become
time-ordered expectation values that we can identify with the
one- and two-particle Green functions $G$ and $G_2$. 
By a subsequent Legendre transform we then can construct a
variational functional in terms of $G$ and $G_2$.  
Let us start out by defining the $n$-body Green function
as
\bea
\lefteqn{ G_n (1 \ldots n , 1' \dots n') = } \nonumber \\
&& (-i)^n \la T_C [ \psia_H(1) \dots \psia_H(n) \psic_H (1')
\ldots \psic_H (n')] \ra
\label{eq:gndef} 
\eea
where we introduced the short notation $1=( \bx_1 t_1)$ and 
where we defined the expectation value of a Heisenberg operator as
\be
\langle \hO \rangle = \frac{\Tr \left\{ \hU (t_0-i\beta,t_0) \hO_H (t) \right\} }
{ \Tr \left\{ \hU (t_0-i\beta,t_0 ) \right\} } \, .
\ee
The many-body Green functions satisfy the following hierarchy 
equations~\cite{MartinSchwinger:PR59,RungeGrossHeinonen:book}
which connect the $n$-body Green function
to the $n+1$ and $n-1$ body Green function:
\bea
\lefteqn{ (i\prt_{t_1} - h (1)) G_n (1 \ldots n , 1' \ldots n') =}  \nonumber \\
&& \sum_{j=1}^n \delta (1 j')(-1)^{n-j} G_{n-1} (2 \ldots n , 1' \dots j'-1 , j'+1 \dots n') \nonumber \\
&& - i \int d\bx v(\bx_1 , \bx) G_{n+1} (1 \ldots n, \bx t_1 , \bx t_1^+, 1' \dots n') .
\label{eq:hierarchy}
\eea
These equations follow directly from the definition of the Green functions, 
the anti-commutation relations of the field operators and the equations of motion of
the evolution operators in Eq.(\ref{eq:umot1}) and (\ref{eq:umot2}).
The Green functions are defined for time-arguments on the 
time contour. Such contour Green functions were first introduced by 
Keldysh~\cite{Keldysh:JETP65}
and are often denoted as Keldysh Green 
functions~\cite{Danielewicz:AP84,Bonitz:book,condmat}
and play an important role in nonequilibrium systems. 
The one-particle Green function $G_1=G$ obeys the boundary 
condition $G(\bx_1 t_0,2)=-G(\bx_1t_0-i\beta,t_2)$ as is readily derived 
using the cyclic property of the trace.
The property $G(1,\bx_2 t_0)=-G(1,\bx_2 t_0-i\beta)$ for the other argument is
likewise easily verified as well as similar relations
for the $n$-body Green functions.
These boundary conditions are sometimes
referred to as the Kubo-Martin-Schwinger 
conditions~\cite{Kubo:JPSJ57,MartinSchwinger:PR59}
and are essential in solving the equations of motion
for the Green function~\cite{condmat}. 
After these preliminaries we are now ready to derive the equations
that connect the one- and two-body Green functions which we will
use to construct the variational functional $Y$.

\section{Relation between the four-point vertex and the Green function}
\label{sec:hedin}

In order to derive a variational energy functional 
in terms of the Green function $G$ and the renormalized 
four-point vertex $\Gamma$ we start out by deriving coupled equations
between these quantities, similar to the familiar
Hedin equations~\cite{Hedin:PR65}.  However, instead of the
usual coupled equations in terms of the Green function $G$ and the
screened interaction $W$ we have equations in terms of
the Green function $G$ and the four-point vertex $\Gamma$. 
Since our aim is to derive equations in terms of
the renormalized four-point vertex it is advantageous to write our
equations in terms of the bare four-point vertex first.
This is most conveniently done within the Hugenholtz
diagram technique~\cite{Hugenholtz:57,Nozieres:book, NegeleOrland:book,BlaizotRipka:book}.
We will therefore first rewrite the two-particle interaction
as a fourpoint function as
\bea
\hat{V} &=&  \half \int d\bx d\bx' v(\br, \brp) \hpsid (\bx)
\hpsid (\bx') \hpsi (\bx') \hpsi (\bx) \nonumber \\
 &=& \frac{1}{4} \int d(\bx_1 \bx_2 \bx_3 \bx_4) V_0(\bx_1 \bx_2 \bx_3 \bx_4) \nonumber \\
&& \times \hpsid (\bx_1) \hpsid (\bx_2 ) \hpsi (\bx_3) \hpsi (\bx_4)  
\eea
where we defined
\bea
V_0(\bx_1 \bx_2 \bx_3 \bx_4) &=& v(\br_1, \br_2) [ \delta (\bx_2 -\bx_3) \delta (\bx_1 -\bx_4) 
\nonumber \\
 && - \delta (\bx_1 -\bx_3) \delta (\bx_2 -\bx_4) ]
\label{eq:V0def}
\eea
\begin{figure}
\begin{center}
\includegraphics[width=7.0cm]{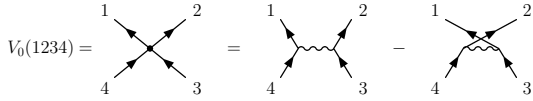}
\end{center}
\caption{Four-point vertex corresponding to the Hugenholtz diagram technique.}
\label{fig:hugvertex}
\end{figure}
This term is used as a basic entity in the Hugenholtz diagram technique and is displayed
pictorially in fig.\ref{fig:hugvertex}.
We now make use of the fact that the Green function
can be obtained as a derivative of the functional
\be
i Y[u] = -\ln \Tr \left\{ U[u](t_0 -i\beta, t_0) \right\}
\ee
with respect to a nonlocal (in space and time) potential $u(12)$,
where
\bea
\lefteqn{ \hU [u] (t_0-i\beta, t_0) = 
T_C \exp (-i \int dt\hH (t) - } \nonumber \\
&& i   \int d1 \int d2 \, \psic (\bx_1) u(1 2) \psia  (\bx_2) ) \, .
\eea
Since this expression contains a double time-integral one has to
define precisely how the time-ordering in this equation is defined. 
This leads to often overlooked subtleties the
details of which are presented in Appendix \ref{genfunc}
where we further show that
\be
G(12) =  i \frac{\delta Y[u]}{\delta u(21)} \, .
\ee
By a subsequent differentiation (see Appendix \ref{genfunc} ) we can 
obtain the two-particle Green function as
\bea
G_2 (1234) = -\frac{\delta G(14)}{\delta u(32)} +
G(14) G(23) .
\label{eq:Gderiv}
\eea
If the derivatives are taken at $u=0$ we obtain the Green functions as 
defined in Eq.(\ref{eq:gndef}). If the derivative is taken at nonzero $u$
then there is no direct relation between the Green function and expectation values
of time-ordered field operators. However, as shown in the Appendix \ref{genfunc}
the Green functions in the presence of a nonlocal potential $u$ still 
satisfy a set of hierarchy equations.
The first ones are
\bea
\lefteqn{ (i \prt_{t_1}- h(1)) G(11') = \delta (11') } \nonumber \\ 
&& + \int d2 ( u(12) + \Sigma (12)) G(21') 
\label{eq:sigmadef} \\
\lefteqn{ (-i \prt_{t_1'}- h(1')) G(11') = \delta (11') } \nonumber \\ 
&& + \int d2 G(12) ( u(21') + \tilde{\Sigma} (21')) .
\label{eq:sigmadefa} 
\eea
where we defined the
self-energy operator $\Sigma$ and its adjoint $\tilde{\Sigma}$ by the equations
\bea
\lefteqn{ \int d2 \, \Sigma (12) G(21') = } \nonumber \\  
&& -\frac{i}{2} \int d(234) \, V(1234) G_2 (4321') 
\label{eq:selfdef1} \\
\lefteqn{ \int d2 \, G(12) \tilde{\Sigma} (21') =} \nonumber \\ 
&& -\frac{i}{2} \int d(234)  \, G_2 (1234) V(4321') 
\label{eq:selfdef2}
\eea
Here we defined 
\bea
V(1234) &=& v(\br_1,\br_2) \delta (t_1,t_2) [ \delta (23) \delta (14) \nonumber \\
&& - \delta (13) \delta (24) ] \theta_{1234}
\label{eq:Vtime}
\eea
where $\delta(ij) =\delta(t_i,t_j) \delta (\bx_i-\bx_j)$ and 
$\theta_{1234}=1$ if $t_1 > t_2 > t_3 > t_4$ (on the contour) and zero otherwise.
The function $\theta_{1234}$ therefore ensures that the operators
have the proper time-ordering before the equal time limits,
described by the delta functions, are taken.
In the next section we will also allow for more general forms of 
$V(1234)$ in order to obtain the two-particle Green function as a
functional derivative with respect to $V$. 
In order to derive a self-consistent set of equations we have to give
a relation between the two-particle Green function and the 
self-energy. We first note that $\Sigma=\tilde{\Sigma}$. This 
can derived by applying to Eq.(\ref{eq:selfdef2}) the operator
$(i \partial_{t_1} -h(1))$ and to Eq.(\ref{eq:selfdef1}) the operator
$(-i\partial_{t_1'} -h(1'))$. With the use of the equations of motion
of the one- and two-particle Green functions from Eq.(\ref{eq:hierarchy}) 
the result then follows.
As a remark we note that for more general initial conditions
$\Sigma$ is not longer equal to $\tilde{\Sigma}$~\cite{bonitzetal:PRE99}. 
From the equality of $\Sigma$ and $\tilde{\Sigma}$
it follows that the Green function has a unique inverse given by
the Dyson equation
\bea
G^{-1} (12) &=& (i \partial_{t_1} -h(1)) \delta (12)- u(12) - \Sigma (12) 
\nonumber \\
 &=& G_0^{-1}(12) -u(12) -\Sigma (12)
\label{eq:Ginverse}
\eea
which satisfies 
\be
\int d2 G^{-1} (12) G(21) = \int d2 G(12) G^{-1} (21') = \delta (11')
\label{eq:inv}
\ee
\begin{figure}
\begin{center}
\includegraphics[width=7.0cm]{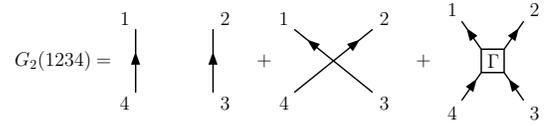}
\end{center}
\caption{Definition of the renormalized four-point vertex $\Gamma$}
\label{fig:four-point vertex}
\end{figure}
For later reference we also defined the inverse $G_0^{-1}$ of the noninteracting
Green function in Eq.(\ref{eq:Ginverse}).
We now define the renormalized four-point vertex by the equation
\bea
 G_2 (1234) = G(14) G(23) - G(13) G(24) \nonumber \\
+ \int d(5678) G(15) G(27) \Gamma (5786) G(83) G (64)
\label{eq:G2Gamma}
\eea
This expression is displayed pictorially in Fig.\ref{fig:four-point vertex}.
The four-point vertex $\Gamma$ has the interpretation of a renormalized 
interaction that describes the scattering of two
particles and will play an important role in our energy functional later.
\begin{figure}
\begin{center}
\includegraphics[width=6.0cm]{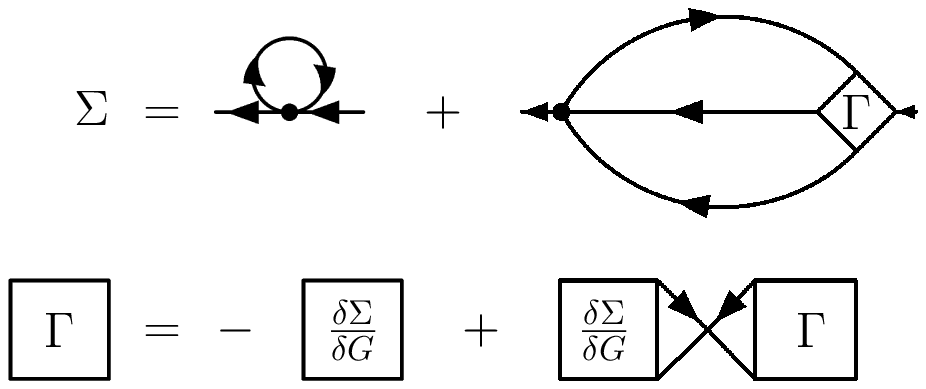}
\end{center}
\caption{Graphical display of the equations
that relate the selfenergy $\Sigma$ to the vertex $\Gamma$.}
\label{fig:hedin}
\end{figure}
By differentiating Eq.(\ref{eq:inv}) with respect to the nonlocal potential $u$
and using Eqs.(\ref{eq:Gderiv}), (\ref{eq:selfdef1}) and (\ref{eq:Ginverse}) 
we can readily derive (along the lines of reference~\cite{Hedin:PR65}) the equations
\bea
\Sigma (18) &=& - i \int d(23) V(1238) G(32) \nonumber  \\
&-& \frac{i}{2} \int d(234567) V(1234) G(36) \nonumber 
\\
&\times&  G(45) 
\Gamma (5678) G(72)
\label{eq:SigmaSC}
\eea
and
\bea
\lefteqn{ \Gamma (1234) = - \frac{\delta \Sigma (14)}{\delta G(32)} +} \nonumber \\
 && \int d(5678) \frac{\delta \Sigma (14)}{\delta G(65)}
G(67) G(85) \Gamma (7238)
\label{eq:GammaSC}
\eea
The Eqs.(\ref{eq:SigmaSC}) and (\ref{eq:GammaSC}),
which are
pictorially displayed in Fig.\ref{fig:hedin}, represent a self-consistent
set of equations that generate the perturbation series for the self-energy 
$\Sigma [G,V]$ in terms of the Green function and the interaction $V$.
For instance,  if one starts by taking $\Gamma=0$ in Eq.(\ref{eq:SigmaSC})
then from Eq.(\ref{eq:GammaSC}) one obtains an improved four-point vertex $\Gamma$
which inserted in Eq.(\ref{eq:SigmaSC}) leads to an improved self-energy.
It should be noted that at a correlated level the simultaneous solution
of these equations is in general computationally demanding~\cite{BickersWhite:PRB91}.
Moreover the self-consistent vertex and self-energy are not related
by a Ward-identity~\cite{Baym:PR62,Varenna:book} although the
response functions calculated from the nonequilibrium Green function in
external fields still satisfy the $f$-sumrule~\cite{KwongBonitz:PRL00,Varenna:book}.
However, as explained in the introduction, our aim is to
obtain total energies. This can be done without solving the computationally
demanding Eqs.(\ref{eq:SigmaSC}) and (\ref{eq:GammaSC}) by using
approximate Green functions and four-point vertices in the
variational functional that we will construct in the next section.

\section{Construction of a variational functional}

\label{sec:varfunc}

In this section we will construct a variational energy or action 
functional of the dressed Green function
$G$ and the renormalized four-point vertex $\Gamma$. The main reason for investigating
such a functional is to obtain in a simple way contributions to the total energy that
correspond to the infinite summation of ladder-type diagrams. 
Such diagrams correspond to an infinite number of terms in the $\Phi$ or
$\Psi$-functional. In the new variables $G$ and $\Gamma$ we have a corresponding functional
$\Xi [G,\Gamma]$ . 
In order to derive the $\Xi$-functional, which we will denote as
the De Dominicis 
functional~\cite{DeDominicis:JMP62,DeDominicis:JMP63,DeDominicisMartin:JMP(1)64,DeDominicisMartin:JMP(2)64,Bloch:book65}, 
we start with the action functional
\be
i Y[ u, V ] = - \ln \Tr \left\{ U[u,V] (t_0 -i\beta, t_0) \right\} 
\ee
which we will regard as a functional of $u$ and $V$, where we
defined
\bea
\lefteqn{ \hU [u,V] (t_0-i\beta, t_0) = 
T_C \exp (-i \int dt\hH_0 (t)  } \nonumber  \\ 
&& -i  \int d1 \int d2 \, \psic (\bx_1) u(1 2) \psia  (\bx_2) \nonumber \\
&&- \frac{i}{4} \int d(1234) V(1234) \nonumber \\
&&  \times  \hpsid (\bx_1) \hpsid (\bx_2 ) \hpsi (\bx_3) \hpsi (\bx_4)  )
\label{eq:UVdef}
\eea
Here $V(1234)$ is a general time-dependent two-body interaction which
we require to have the following symmetry properties
\be
V(1234)=-V(2134)=-V(1243)=V(2143)
\label{eq:symmetry}
\ee
This will guarantee that the Feynman rules of the Hugenholtz diagram
method are satisfied.
Eventually, when we have derived our equations, we will set $V$
equal to expression $V_0$ of (\ref{eq:Vtime}). To give precise meaning
to expression Eq.(\ref{eq:UVdef}) we again have to specify how
the time-ordering is defined when we expand the exponent. This
is done in Appendix \ref{genfunc2}
where we show that
\bea
i \frac{\delta Y}{\delta u(2 1) } &=& G(1 2) 
\label{eq:G1derv}\\
i\frac{\delta Y}{\delta V(4 3 2 1) } &=& -\frac{i}{4} G_2 (1 2 3 4)
\label{eq:G2derv}
\eea
In Appendix \ref{genfunc2} it is further demonstrated that these
one- and two-particle Green functions, even in the presence of time nonlocal
fields, are related by the first equation of
motion of the Martin-Schwinger hierarchy.
By a Legendre transform we can now  construct a functional of $G$ and $G_2$
\bea
\lefteqn{  F[ G, G_2 ] = i Y [ u[G,G_2], V[G,G_2] ] }\nonumber \\
 && - \int d(12) u(21) G(12) \nonumber \\
&& + \frac{i}{4} \int d(1234) V(4321) G_2 (1234)
\eea
where we now regard $u$ and $V$ as functionals of $G$ and $G_2$.
This functional satisfies
\bea
 \frac{\delta F}{\delta G(12)} &=& -u(21) \\
 \frac{\delta F}{\delta G_2(1234)} &=& \frac{i}{4} V(4321) .
\eea
Therefore the functional
\bea
\lefteqn{ i Y[ G, G_2 ] =  F[ G, G_2 ] } \\
&& + \int d(12) u(21) G(12) \nonumber \\
&& - \frac{i}{4} \int d(1234) V(4321) G_2 (1234)
\label{eq:S}
\eea
for {\em fixed}  $u$ and $V$ is a stationary functional of $G$ and $G_2$, i.e.
\bea
i \frac{\delta Y}{\delta G(12)} &=& 0 \\
i \frac{\delta Y}{\delta G_2(1234)} &=& 0
\eea
where we will eventually be interested in the case $u=0$ and $V=V_0$. 
We can now modify the
functional $Y[G,G_2]$ such that, rather than the two-particle Green function,
we can use the renormalized four-point vertex $\Gamma$ as a basic variable.
For this purpose we use Eq.(\ref{eq:G2Gamma})
which is displayed pictorially in 
fig.\ref{fig:four-point vertex} and which 
gives $G_2 [G,\Gamma]$ as an explicit functional
of $G$ and $\Gamma$.
We then define the functional
\be
H[ G, \Gamma ] = F [ G, G_2 [G, \Gamma] ]
\ee
which is a functional of the Green function $G$ and the
four-point vertex $\Gamma$.
Then for fixed $\Gamma$ we have
\bea
\frac{\delta H}{\delta G(12)} &=& \frac{\delta F}{\delta G(12)} +
\int d(3456) \, \frac{\delta F}{\delta G_2 (3456)} 
\frac{\delta G_2 (3456)}{\delta G(12)} \nonumber \\
&=& -u(21) - \Sigma (21) -\tilde{\Sigma}_C (21)
\eea
where we defined
\bea
\Sigma (12) &=& \Sigma^{HF} (12) + \Sigma_C (12) \\
\Sigma^{HF} (14) &=& -i \int d(23) V(1234) G(32) 
\label{eq:self1}  \\
\Sigma_C (18) &=&  -\frac{i}{2} \int d(234567)
V(1234) G(36) \nonumber \\ && \times G(45) \Gamma (5678) G(72) 
\label{eq:self2}  \\
\tilde{\Sigma}_C (18) &=& -\frac{i}{2} \int d(234567)
\Gamma (1234) G(36) \nonumber \\ && \times G(45) V (5678) G(72) 
\label{eq:selfenergy}
\eea
From Eq.(\ref{eq:SigmaSC}) we recognize these terms as selfenergy diagrams.
\begin{figure}
\begin{center}
\includegraphics[width=6.0cm]{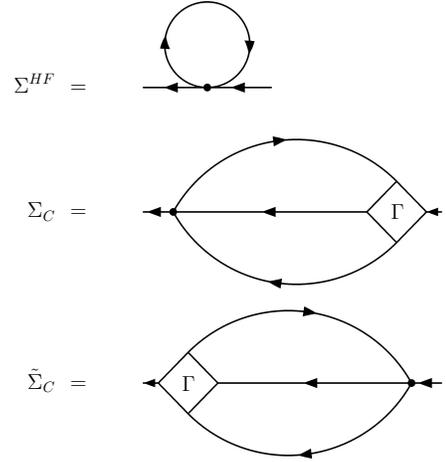}
\end{center}
\caption{Graphical display of the self-energy terms. 
The small dot denotes the bare vertex $V$
and the big square denotes the full four-point vertex $\Gamma$.}
\label{fig:selfener}
\end{figure}
They are displayed graphically in Fig. \ref{fig:selfener}.
We recognize the first term in Eq.(\ref{eq:self1}) for $V=V_0$
as the Hartree-Fock part of the self-energy.
The second part $\Sigma_C$ of Eq.(\ref{eq:self2}) involving the
four-point vertex $\Gamma$ describes the time-nonlocal correlation part
of the self-energy.
The third part $\tilde{\Sigma}_C$ on Eq.(\ref{eq:selfenergy}) 
is the adjoint correlation part of the 
self-energy. As mentioned earlier we
can show from the Kubo-Martin-Schwinger boundary conditions for
a system initially in thermodynamic equilibrium that
$\tilde{\Sigma}_C (1 2) = \Sigma_C (1 2)$. However, in the following we will keep the
tilde on the self-energy to keep track of the origin of this term.
For fixed $G$ we can also calculate the derivative with respect to $\Gamma$
for which we have
\bea
 \frac{\delta H}{\delta \Gamma(1234)} 
&=& \int d(5678) \frac{\delta F}{\delta G_2 (5678)}
\frac{\delta G_2 (5678)}{\delta \Gamma (1234)} \nonumber \\
&=& \frac{i}{4} \tilde{V}(4321)
\eea
where we defined 
\be
\tilde{V} (1234) = \int d(5678) G(15) G(26) V(5678) G(73) G(84)
\label{eq:tilde}
\ee
which is simply a bare vertex dressed with two ingoing and
two outgoing dressed Green function lines.
Using the functional $H$ we can now regard the expression $iY$ of Eq.(\ref{eq:S}) 
as a functional of $G$ and $\Gamma$, i.e.
\bea
\lefteqn{ i Y[ G, \Gamma] = H [ G, \Gamma ] } \nonumber \\
&& + \int d(12) \, u(21) G(12) \nonumber \\
&& - \frac{i}{4} \int d(1234) V(4321) G_2 [G,\Gamma] (1234)
\label{eq:H1}
\eea
which is a stationary functional of $G$ and $\Gamma$ for fixed $u$ and $V$. 
We have thus achieved our first goal and expressed the action $iY$ as
functional of $G$ and $\Gamma$.
Our next step is
to specify the functional $H$ in more detail. 
The variations in $H$ are given by the expression
\bea
\delta H &=&  \int d(12) (-u(21)-\Sigma (21)-\tilde{\Sigma}_C (21) ) \delta G(12) \nonumber \\
 && +
\frac{i}{4} \int d(1234) \tilde{V}(4321) \delta \Gamma (1234) \nonumber \\
&=& \int d(12) (G^{-1}(21) - G_0^{-1}(21) -\tilde{\Sigma}_C (21) )  \delta G(12) \nonumber \\
&& +
\frac{i}{4} \int d(1234) \tilde{V}(4321) \delta \Gamma (1234)
\label{eq:Hderv}
\eea
and hence we see that it is
convenient to split up $H$ as follows
\bea
H[ G, \Gamma ] &=& -\tr \left\{ \ln (- G^{-1} ) \right\} \nonumber \\
 && -  \tr \left\{ G_0^{-1} (G-G_0) \right\} - \Xi [G,\Gamma ] 
\label{eq:Hdef}
\eea
This equation defines a new functional $\Xi [G, \Gamma ]$ which will be the
central object for the rest of the paper.
In Eq.(\ref{eq:Hdef}) we further defined the trace $\tr$ (not to be confused with the
thermodynamic trace $\Tr$) as
\be
\tr A = \int d1 \, A(1,1^+)
\ee
where $1^+$ denotes that time $t_1$ is approached from above on the contour.
The definition of the $\Xi$-functional in Eq.(\ref{eq:Hdef})
is convenient because then we have
\bea
\frac{\delta H}{\delta G(12)} &=& G^{-1}(21) - G_0^{-1}(21) - 
\frac{\delta \Xi}{\delta G(12)}  \\
\frac{\delta H}{\delta \Gamma (1234)} &=& -\frac{\delta \Xi}{\delta \Gamma (1234)}
\eea
and therefore from Eq.(\ref{eq:Hderv}) we see that the functional $\Xi [G,\Gamma ]$ 
satisfies the equations
\bea
\frac{\delta \Xi}{\delta \Gamma (1234)} &=& -\frac{i}{4} \tilde{V} (4321)
\label{eq:Yderiv} \\
\frac{\delta \Xi}{\delta G(12)} &=& \tilde{\Sigma}_C (21)
\label{eq:Xideriv2}
\eea
\begin{figure}
\begin{center}
\includegraphics[width=7.0cm]{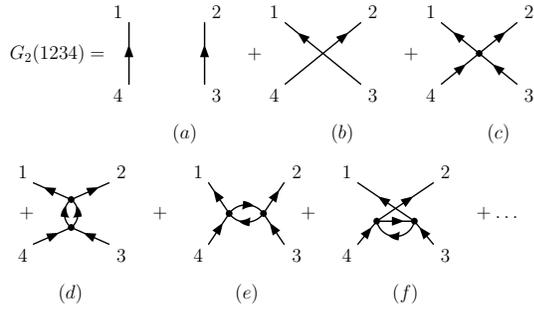}
\end{center}
\caption{Expansion of the 2-particle Green function $G_2$ in terms of the full $G$.
The dot denotes the bare vertex $V$.}
\label{fig:G2}
\end{figure}
The functional $\Xi$ is therefore directly related to the correlation
part of the self-energy. To describe the correlations in the
system it is therefore necessary to further study the structure
of the $\Xi$-functional, which we will do in detail in the next section.

Note that in Eq.(\ref{eq:Hdef}) we could also have written $\ln(G^{-1})$
rather than $\ln (-G^{-1})$. These terms differ only by a (possibly infinite)
constant and depend on the definition of the branch cut of the
logarithm. However, the particular definition here
reduces properly to the grand potential 
of the noninteracting system when the interactions are 
switched off~\cite{LuttingerWard:PR60}.
The final De Dominicis functional (for $u=0$ ) is thus given 
from Eq.(\ref{eq:H1}) and (\ref{eq:Hdef}) by
\bea
\lefteqn{  iY[ G, \Gamma ] =  -\tr \left\{ \ln (-G^{-1}) \right\} 
-  \tr \left\{ G_0^{-1} (G-G_0) \right\} }
\nonumber \\
&&  -  \Xi [ G, \Gamma ] \nonumber \\
&& - \frac{i}{4}  \int d(1234) V_0 (1234) G_2 [ G, \Gamma] (4321)
\label{eq:func1}
\eea
We can check that in the absence of interactions we have 
$iY = -\tr \ln \{ -G_0^{-1} \}$ which yields the grand potential
of the noninteracting system, as we will discuss in more detail later.
Let us now check the variational property.
The derivatives of $iY$ with respect to $G$ and
$\Gamma$ are given by
\bea
i\frac{\delta Y}{\delta \Gamma (1234)} &=& 
\frac{i}{4} (\tilde{V} (4321) - \tilde{V}_0 (4321))  
\label{eq:varia1} \\
i\frac{\delta Y}{\delta G(12)} &=& G^{-1} (21) - G_0^{-1} (21) 
 -\tilde{\Sigma}_C (21; V) \nonumber \\
&& + \Sigma (21;V_0) + \tilde{\Sigma}_C (21;V_0)
\label{eq:varia2}
\eea
where we used that
\bea
- \frac{i}{4} \frac{\delta}{\delta G (56)}  
  \int d(1234) V_0 (1234) G_2 [ G, \Gamma] (4321) \nonumber \\
= \Sigma (65;V_0) + \tilde{\Sigma}_C (65;V_0)
\eea
The variational equations that are obtained by putting the
derivatives (\ref{eq:varia1})
and (\ref{eq:varia2}) equal to zero, are obviously solved
for the $G$ and $\Gamma$ that self-consistently solve the
equations
\bea
G^{-1} &=& G_0^{-1} - \Sigma [G,\Gamma] \\
\tilde{V}_0 &=& \tilde{V} [G, \Gamma] 
\label{eq:Vfunc}
\eea
where $\Sigma$ is calculated from Eqs.(\ref{eq:self1}) and Eq.(\ref{eq:self2}).
Therefore the functional $Y[G,\Gamma]$ is stationary whenever the
Dyson equation is obeyed and whenever the electron-electron interaction expanded
in $G$ and $\Gamma$ is equal to the specified interaction $V_0$.
Equation (\ref{eq:func1}) for the variational functional $Y[G,\Gamma]$ 
is the first basic result of this work. However, before it can be used in 
actual calculations we have, among others, to specify the specific
structure of the functional $\Xi [G, \Gamma]$. We will show that for several infinite series of
diagrammatic terms
contributing to this functional we can find explicit
expressions in terms of $G$ and $\Gamma$. 
To do this we first have to study the functional $\tilde{V}[G,\Gamma]$ 
of Eq.(\ref{eq:Vfunc}). This is the topic of the next section.

\section{Structure of the $\Xi$-functional}

\begin{figure}
\begin{center}
\includegraphics[width=7.0cm]{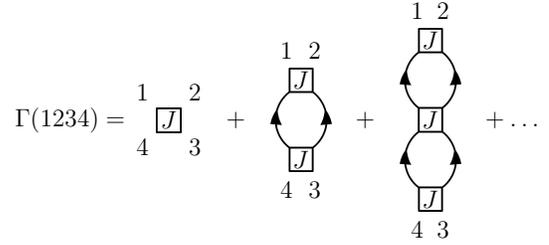}
\end{center}
\caption{Expression of $\Gamma$ in terms of $J$-blocks}
\label{fig:cpp}
\end{figure}

\begin{figure}
\begin{center}
\includegraphics[width=7.0cm]{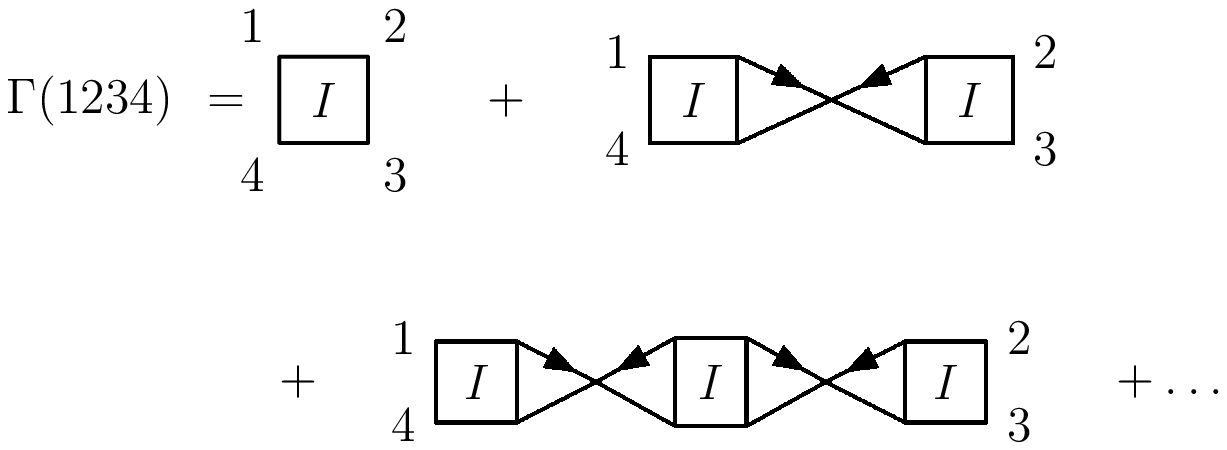}
\end{center}
\caption{Expression of $\Gamma$ in terms of $I$-blocks}
\label{fig:cph}
\end{figure}

\begin{figure}
\begin{center}
\includegraphics[width=7.0cm]{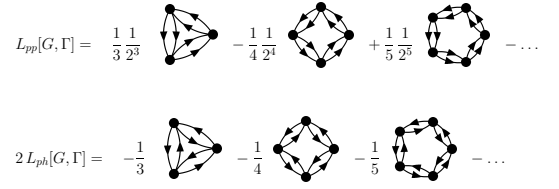}
\end{center}
\caption{Expansion of the functionals $L_{pp}$ and $L_{ph}$ in diagrams.
The four-point vertex $\Gamma$ is denoted with a big black dot.}
\label{fig:logs}
\end{figure}

\begin{figure}
\begin{center}
\includegraphics[width=7.0cm]{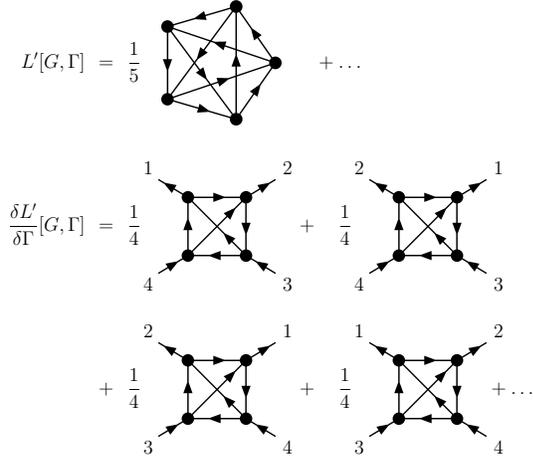}
\end{center}
\caption{The first term in the expansion of the $L'$-functional and its
functional derivative with respect to $\Gamma$. For clarity in drawing the
diagrams for $\delta L'/\delta \Gamma$ we interchanged the endpoint labels
rather than making the outgoing lines cross.}
\label{fig:penta}
\end{figure}

In this section we analyze in more detail the 
diagramatic structure of the four-point vertex $\Gamma$ and the
functional $\tilde{V}[G,\Gamma]$ of Eq.(\ref{eq:Vfunc})
which will allow us to obtain more explicit expressions
for the functional $\Xi$. These quantities can
be directly obtained from a diagrammatic expansion of the
two-particle Green function. 
If we express the diagrams in terms of the fully dressed
Green function $G$ we only need to
consider diagrams that do not contain any self-energy insertions.
Since different authors use different definitions and drawing conventions
for the two-particle Green function, it is important to be clear
about them. We strictly follow the sign, loop rule and
drawing conventions of reference~\cite{RungeGrossHeinonen:book}
with the small difference that we use 
Hugenholtz diagrams~\cite{Hugenholtz:57,Nozieres:book,NegeleOrland:book,BlaizotRipka:book}. For clarity our
Feynman rules are given in Appendix~\ref{section:frules}.
In fig. \ref{fig:G2} we show the first and second order 
Hugenholtz diagrams
in terms of the fully dressed Green function $G$ that 
contribute to the two-particle Green function $G_2$.
We see that we can write $\Gamma$ as a sum of four classes of diagrams.
There are three classes of the form $(ab,cd)$ which denote diagrams 
which by removal of two internal Green function
lines can separate the diagram into two parts, one part being connected to
the external points $ab$ and one part being connected to points $cd$. 
The class $(12,34)$ contains diagrams of the particle-particle type, such as
diagram $(d)$ in fig.\ref{fig:G2},
and will be denoted by $\Gamma_{pp}$. There are two classes of
particle-hole type, namely $(14,32)$ and $(13,24)$ which will be denoted
by $\Gamma_{ph}^A$ and $\Gamma_{ph}^B$. Examples of diagrams of these types are diagrams
$(e)$ and $(f)$ in fig.\ref{fig:G2}. 
The remaining diagrams which do not fall into one of these classes are
denoted by $\Gamma_0$ (such as diagram $(c)$ in fig.\ref{fig:G2}).
We can therefore write
\bea
\Gamma(1234) &=& \Gamma_{pp} (1234) + \Gamma_{ph}^A (1234) \nonumber \\
&& + \Gamma_{ph}^B(1234) + \Gamma_0 (1234)
\label{eq:splitup}
\eea
The simplest diagram in class $\Gamma_0$ is simply the bare vertex
$iV(1234)$ (i.e.diagram $(c)$ in fig.\ref{fig:G2}, the factor $i$ follows
from the Feynman rules in Appendix \ref{section:frules}). 
Since this diagram is special we separate it off from $\Gamma_0$
and define the remaining diagrams $\Gamma_0'$ by the equation
\be
\Gamma_0 (1234) = \Gamma_0' (1234) + iV(1234)
\ee
Using Eq.(\ref{eq:splitup}) we can then write
\bea
\lefteqn{ -iV(1234) = \Gamma_{pp} (1234) + \Gamma_{ph}^A (1234) } \nonumber \\
&& + \Gamma_{ph}^B(1234) + \Gamma_0' (1234)
- \Gamma(1234)
\label{eq:splitup2}
\eea
We will now first show how all the terms on the right-hand side of
this equation can be constructed as
a functional of $\Gamma$. When we have done this we can insert this functional
into Eq.(\ref{eq:Yderiv}) and perform the integration with respect to $\Gamma$ and thereby
construct our desired functional $\Xi [ G,\Gamma]$. \\
Let us start with the
particle-particle diagrams $\Gamma_{pp}$.
The contribution of all diagrams for $\Gamma_{pp}$ can be written as sums of
blocks of diagrams $J$ connected with two parallel Green function lines
(see fig.\ref{fig:cpp} ). Each of these blocks $J$ contains diagrams which cannot disconnect
points $(12)$ and $(34)$ by cutting two Green function lines
(such blocks are called simple with respect to $(12)$ and $(34)$
in the terminology of De Dominicis)
and therefore each $J$-block does not contain diagrams of the type $\Gamma_{pp}$. 
We thus have
\be
J(1234) = \Gamma (1234)- \Gamma_{pp}(1234)
\label{eq:GammaJ1}
\ee 
We introduce a convenient matrix notation
\bea
 \langle 12 | J | 34 \rangle  &=& J(1234) \\
\langle 12 | GG | 34 \rangle &=& G(1 3 ) G(2 4) 
\eea
Within this notation we can, for instance,
conveniently write $C=AB$ instead of
\bea
C(1234) &=& \langle 12 | AB | 34 \rangle \nonumber \\
&=& \int d(56) 
\langle 12| A | 56 \rangle \langle 56 | B | 34 \rangle \nonumber \\
&=& \int d(56) A(1256) B(5634)
\eea
If we use this notation, 
then from the Feynman rules in Appendix \ref{section:frules} one
can readily convince oneself that in matrix notation we simply have
\bea
\Gamma &=& J + \Gamma_{pp} \nonumber \\
&=& J + \frac{1}{2} J GG J + (\frac{1}{2})^2 J GG J GG J + \ldots \nonumber \\
&=&  J + \frac{1}{2} J GG \Gamma
\label{eq:GammaJ} 
\eea
where for every pair of Green function lines we have to add a
factor of $\frac{1}{2}$ (see \cite{Nozieres:book,NegeleOrland:book,BlaizotRipka:book,Bloch:book65}).
This follows because for any diagram contributing to $J$, the diagram with
outgoing lines interchanged leads to the same diagram 
for $\Gamma$
(for the simple diagram $iV$ in $J$ it follows from Feynman rule 5
in Appendix \ref{section:frules}). 
Relation Eq.(\ref{eq:GammaJ}) 
allows us to express $\Gamma_{pp}$ in terms of $\Gamma$.
We have 
\be
J = \Gamma (1 + \frac{1}{2} GG \Gamma )^{-1} 
\ee
In combination with Eq.(\ref{eq:GammaJ1}) this then gives
\be
\Gamma_{pp} = \Gamma - \Gamma (1 + \frac{1}{2}  GG \Gamma )^{-1} 
\label{eq:Cpp}
\ee
which expresses
$\Gamma_{pp}$ in terms of $\Gamma$. 
Let us now do the same for the particle-hole diagrams.
Since
\be
\Gamma_{ph}^B (1234) = -\Gamma_{ph}^A (2134)
\ee
we only need to construct $\Gamma_{ph}^A$ as a functional of $\Gamma$.
For the particle-hole diagrams $\Gamma_{ph}^A$ we can follow a
similar reasoning as for $\Gamma_{pp}$ and
we first write $\Gamma$ in terms
of repeated blocks $I$ given by
\be
I (1234) = \Gamma(1234)- \Gamma_{ph}^A (1234) .
\ee
The expression for $\Gamma$ in terms of $I$ is displayed in fig.\ref{fig:cph}.
If we use the notation
\bea
 \langle 4 1 | \bar{I} | 2 3 \rangle  &=& I(1234) \\
\langle 12 | \widehat{GG} | 34 \rangle &=&  G(3 1) G(2 4) 
\eea
where in the first term we defined a new matrix $\bar{I}$ by a cyclic
permutation of the indices,
then (again using the Feynman rules of Appendix \ref{section:frules}) 
we have in matrix notation 
\bea
\bar{\Gamma} &=& \bar{I} + \bar{\Gamma}_{ph}^A \nonumber \\
&=& \bar{I} - \bar{I} \widehat{GG} \bar{I} + \bar{I} \widehat{GG} \bar{I}
\widehat{GG} \bar{I} - \dots \nonumber \\
&=& \bar{I} - \bar{I} \widehat{GG}  \bar{\Gamma}
\label{eq:GammaI}
\eea
where the alternating signs in Eq.(\ref{eq:GammaI}) are related to
Feynman rule 4 in Appendix \ref{section:frules}.
As a remark we note that from Eq.(\ref{eq:GammaI}) and (\ref{eq:GammaSC}) we see that there is
a simple relation between $I$ and the self-energy:
\be
I(1234) = - \frac{\delta \Sigma (14)}{\delta G(32)}
\ee
One can indeed check, by iterating equations
(\ref{eq:SigmaSC}) and (\ref{eq:GammaSC}), that the
term $\delta \Sigma/ \delta G$ only yields diagrams that contribute to $I$.
We can now express $\bar{\Gamma}_{ph}^A$ in terms of $\bar{\Gamma}$. We have
\be
\bar{I} = \bar{\Gamma} (1 -  \widehat{GG} \bar{\Gamma} )^{-1} 
\ee 
which gives
\be
\bar{\Gamma}_{ph}^A = \bar{\Gamma} - \bar{\Gamma} 
(1 -  \widehat{GG} \bar{\Gamma})^{-1}
\label{eq:Cph}
\ee
Before discussing the last set of diagrams $\Gamma_0'$
let us see if we can integrate the functionals 
$\Gamma_{pp}$ and $\Gamma_{ph}^A$ that we obtained sofar.
To do this we first make a general remark about functional derivatives.
We consider a given fourpoint function $a [\Gamma] (1234)$
that we want to integrate
with respect to $\Gamma$ to obtain a functional $A$, i.e.
\be
\delta A = \int d(1234) \, a[\Gamma] (1234) \delta \Gamma (1234)
\ee
In our case we want to do this for $a[\Gamma]$ being 
$\Gamma_{pp}$, $\Gamma_{ph}^A$, $\Gamma_{ph}^B$ and $\Gamma_0'$.
Because $\delta \Gamma$ has the symmetry property of Eq.(\ref{eq:symmetry})
this can also be written as
\bea
\delta A &=& \frac{1}{4} \int d(1234) \, [ a(1234) - a(2134) \nonumber  \\
&& + a(3412 ) - a(1243) ] \delta \Gamma(1234)
\label{eq:fourterms}
\eea
Therefore any part of $a$ which is symmetric in the indices $(12)$ or $(34)$ 
(or anti-symmetric with respect to the interchange of pair $(12)$ and pair $(34)$) will
not contribute to this variation. Therefore only certain (anti-)symmetric parts of $a$ are
uniquely determined as functional derivatives.
This does not pose a problem if the functional $a$ we want to integrate
already has the same symmetry as $\Gamma$. This applies for instance to
$\Gamma_{pp}$ and $\Gamma_0'$. However, the function $\Gamma_{ph}^A (1234)$
is not anti-symmetric in the indices $(12)$ and $(34)$.
However, the combination
\be
\Gamma_{ph}^A (1234) - \Gamma_{ph}^A (2134) = \Gamma_{ph}^A (1234) + \Gamma_{ph}^B (1234)
\ee
has this property and therefore
\bea
&& 2 \int d(1234) \, \Gamma_{ph}^A (1234) \delta \Gamma(1234) = \nonumber \\
&& \int d(1234) \, [ \Gamma_{ph}^A (1234) + \Gamma_{ph}^B (1234) ]
\delta \Gamma(1234)
\eea
We can therefore obtain $\Gamma_{ph}^A + \Gamma_{ph}^B$ as a functional
derivative by formally integrating $\Gamma_{ph}^A$ and multiplying the 
resulting functional by $2$.
With this in mind we can now address the integration of $\tilde{V}$
in the right hand side of Eq.(\ref{eq:Yderiv}) with respect to $\Gamma$.
Using Eq.(\ref{eq:splitup2}) we can write
\bea
-i \tilde{V}(1234) &=& \tilde{\Gamma}_{pp} (1234) \nonumber \\
&+& 
\tilde{\Gamma}_{ph}^A (1234) + \tilde{\Gamma}_{ph}^B(1234) \nonumber \\
&+& \tilde{\Gamma}_0' (1234)
- \tilde{\Gamma} (1234)
\eea
where the expressions with the tilde are defined as in Eq.(\ref{eq:tilde}). 
Let us start by integrating $\tilde{\Gamma}_{pp}$ with respect to $\Gamma$.
Using Eq.(\ref{eq:Cpp}) and taking into account the factor $1/4$
in Eq.(\ref{eq:Yderiv}) we have
\bea
\frac{1}{4} \tilde{\Gamma}_{pp} &=& \frac{1}{4} GG \Gamma_{pp} GG \nonumber \\
&=& \frac{1}{4} GG \Gamma [ 1 - (1+\frac{1}{2}  GG \Gamma )^{-1} ]  GG \nonumber \\
&=& \frac{1}{4} GG \Gamma GG  - \frac{1}{2} [ 1 - (1+\frac{1}{2}  GG \Gamma )^{-1} ] GG
\nonumber \\
&=& \frac{\delta L_{pp} [ G,\Gamma]}{\delta \Gamma}
\eea 
where we defined
\bea
L_{pp} [ G, \Gamma ] &=&
\frac{1}{8} \tr \left\{ GG \Gamma GG \Gamma\right\}  
-\frac{1}{2} \tr \left\{ GG \Gamma \right\} \nonumber \\
&& + \tr \left\{ \ln (1 + \frac{1}{2} GG \Gamma ) \right\}
\eea
In this expression the trace $\tr$ (not to be confused with the
thermodynamic trace $\Tr$) for two-particle functions is defined as
\be
\tr \left\{ A \right\} = \int d(1 2) \langle 12 | A | 12 \rangle
\ee
The diagrammatic expansion of the functional $L_{pp}$ 
is displayed in the upper part of fig.\ref{fig:logs}.
Let us now consider the particle-hole diagrams. Since
\bea
 \tr \left\{ \widehat{GG} \bar{\Gamma}_{ph}^A \widehat{GG} \delta \bar{\Gamma} \right\} 
= \tr \left\{ \tilde{\Gamma}_{ph}^A \delta \Gamma \right\}
\eea
it is sufficient to integrate $\widehat{GG} \bar{\Gamma}_{ph}^A \widehat{GG}$ with respect
to $\bar{\Gamma}$. We have using Eq.(\ref{eq:Cph})
\bea
\frac{1}{4} \tilde{\bar{\Gamma}}_{ph}^A 
&=& \frac{1}{4} \widehat{GG} \bar{\Gamma}_{ph}^A \widehat{GG} \nonumber \\
&=& \frac{1}{4} \widehat{GG} \bar{\Gamma} [ 1 - (1-  \widehat{GG} \bar{\Gamma} )^{-1} ]  
\widehat{GG} \nonumber \\
&=& \frac{1}{4}  \widehat{GG} \bar{\Gamma}_{ph}^A \widehat{GG}
 + \frac{1}{4}  [ 1 - (1-  \widehat{GG} \bar{\Gamma} )^{-1} ] \widehat{GG} \nonumber \\
&=& \frac{1}{2} \frac{\delta L_{ph} [ G,\Gamma]}{\delta \bar{\Gamma} }
\label{eq:phderiv}
\eea 
where we defined the functional
\bea
L_{ph} [G, \Gamma ] &=&
\frac{1}{4} \tr \left\{  \widehat{GG} \bar{\Gamma} 
\widehat{GG} \bar{\Gamma} \right\} + \frac{1}{2} \tr \left\{ \widehat{GG} \bar{\Gamma} \right\}
\nonumber \\
&& + \frac{1}{2} \tr \left\{ \ln (1 -   \widehat{GG} \bar{\Gamma} ) \right\}
\eea
The diagrammatic expansion of the functional $L_{ph}$
is displayed in the lower part of fig.\ref{fig:logs}.
Now since
\bea
\tr \left\{ (\tilde{\Gamma}_{ph}^A + \tilde{\Gamma}_{ph}^B) \delta \Gamma  \right\} &=&
2 \,\tr \left\{ \tilde{\Gamma}_{ph}^A  \delta \Gamma  \right\} \nonumber \\
&=&
2 \, \tr \left\{ \widehat{GG} \bar{\Gamma}_{ph}^A \widehat{GG} \delta \bar{\Gamma} \right\}
\eea
we obtain
\be
\frac{1}{4} ( \tilde{\Gamma}_{ph}^A + \tilde{\Gamma}_{ph}^B ) = 
\frac{\delta L_{ph} [ G,\Gamma]}{\delta \Gamma }
\ee
We now collect our results and define
\bea
L[ G, \Gamma ] &=&  L_{pp} [G, \Gamma] + L_{ph} [G, \Gamma] \nonumber \\
&& - \frac{1}{8} \tr \left\{ GG \Gamma GG \Gamma \right \}
\eea
This functional $L$ has the property
\be
\frac{\delta L}{\delta \Gamma} = \frac{1}{4} 
( \tilde{\Gamma}_{pp} + \tilde{\Gamma}_{ph}^A + 
\tilde{\Gamma}_{ph}^B - \tilde{\Gamma})
\ee
Using this functional we can now split up the functional $\Xi$ further as
\be
\Xi [G, \Gamma] = L [G, \Gamma] + L' [G, \Gamma]
\label{eq:Lprimedef}
\ee
This defines a new functional $L' [G,\Gamma]$.
Then from Eq.(\ref{eq:Yderiv}) we see that
if we differentiate both sides of Eq.(\ref{eq:Lprimedef}) with 
respect to $\Gamma$ we obtain
\be
\frac{\delta \Xi}{\delta \Gamma} = -\frac{i}{4} \tilde{V} = \frac{1}{4}
(\tilde{\Gamma}_{pp} + \tilde{\Gamma}_{ph}^A + 
\tilde{\Gamma}_{ph}^B - \tilde{\Gamma}) + \frac{\delta L'}{\delta \Gamma}
\ee
We therefore see by comparing to Eq.(\ref{eq:splitup2})
that the functional $L'$ must satisfy
\be
\frac{1}{4} \tilde{\Gamma}_0' = \frac{\delta L' [ G, \Gamma] }{\delta  \Gamma}
\ee
This functional can not be written out explicitly, but since
$\Gamma_0'$ is well-defined diagrammatically the functional $L'$ does
have a diagrammatic expansion. The first term in this expansion is
displayed in fig.\ref{fig:penta} together with its functional
derivative. Note that the derivative yields four diagrams in accordance
with Eq.(\ref{eq:fourterms}). 
We can further consider the functional derivative of the functional
$\Xi [G,\Gamma]$ with respect to $G$.
According to Eq.(\ref{eq:Xideriv2}) this yields self-energy diagrams, as
is also seen from the diagrammatic expansion of $\Xi$.
The $G$-derivatives of $L_{pp}$, $L_{ph}$ and $L'$ lead to correlation self-energy
diagrams $\Sigma_{C,pp}[G,\Gamma]$, $\Sigma_{C,ph} [G,\Gamma]$
and $\Sigma_C' [G,\Gamma]$ in terms of $G$ and $\Gamma$ that fall into
different topological classes.

We now again collect our results and find from Eq.(\ref{eq:func1}) that
the final De Dominicis functional (for $u=0$ ) is given by
\bea
\lefteqn{ iY[ G, \Gamma ] =  -\tr \left\{ \ln (-G^{-1}) \right\} 
-  \tr \left\{ G_0^{-1} (G-G_0) \right\} }
\nonumber \\
&& - L [ G, \Gamma] - L' [ G, \Gamma ] - 
\frac{i}{4} \tr \left\{ V_0 G_2 [ G, \Gamma] \right\}
\label{eq:Kleinform}
\eea
We finally write the functional in a different form using the Dyson equation
of Eq.(\ref{eq:Ginverse})
\bea
\lefteqn{ iY[G,\Gamma] = -\tr \left\{ \ln ( \Sigma  - G_0^{-1} ) \right\} 
-\tr \left\{ \Sigma G \right\} } \nonumber \\
&& - L [G,\Gamma] - L' [G,\Gamma]  - 
\frac{i}{4} \tr \left\{ V_0 G_2 [ G, \Gamma] \right\}
\label{eq:LWform}
\eea
We can readily check the variationally property of this functional. We then find that
\bea
i \delta Y &=&  -\tr \left\{ [(\Sigma - G_0^{-1} )^{-1} + G ] \delta \Sigma \right\} 
\nonumber \\
&&
- \tr \left\{ [ \Sigma (V) -\Sigma (V_0) + \tilde{\Sigma}_C (V) -
 \tilde{\Sigma}_C (V_0) ] \delta G \right\}
\nonumber \\
&& + \frac{i}{4} \tr \left\{ ( \tilde{V} -\tilde{V}_0 ) \delta \Gamma \right\} =0 
\label{eq:property}
\eea
whenever $V[G,\Gamma ]=V_0$ for a self-consistent solution of the
Dyson equation.  The variational functional
(\ref{eq:LWform}) together with the variational property (\ref{eq:property})
is the central result of this work. In the next sections we will
investigate the practical evaluation of this functional. 
It is important to note that although the functionals in Eq.(\ref{eq:Kleinform})
and (\ref{eq:LWform}) are equivalent when evaluated on the fully self-consistent
$G$ and $\Gamma$ obtained from the Dyson equation and $V[G,\Gamma ]=V_0$
this is not true anymore when evaluated at approximate $G$ and $\Gamma$.
In accordance with ref.\cite{DahlenvonBarth:PRB04} the
functional forms in Eqs.(\ref{eq:Kleinform}) and (\ref{eq:LWform})
will be denoted as the Klein-form and Luttinger-Ward-form of the
functional $Y$. It was demonstrated in the $\Phi$-formalism 
that the Luttinger-Ward form of the functional is
more stable (has a smaller second derivative) when used for the
calculation of total energies~\cite{Dahlenetal:PRA06}. 
We will therefore in the following
use the Luttinger-Ward form of the functional. 

\section{$\Xi$-derivable theories are conserving}

In this section we will show that any approximate $\Xi$-functional leads
to a corresponding $\Phi$-functional. 
Since we know from the work of Baym~\cite{Baym:PR62} that
any $\Phi$-derivable theory is conserving it follows
that also $\Xi$-derivable theories are conserving, i.e.
they respect the macroscopic conservation laws, such
a momentum, energy and particle number conservation
and related constraints such as the virial theorem~\cite{DahlenvanLeeuwen:JCP05}.
Consider any approximate $\Xi$-functional. Then from the
variational equation
\be
\frac{\delta \Xi [G,\Gamma]}{\delta \Gamma } = -\frac{i}{4} \tilde{V_0}
\ee
we can construct $\Gamma [G, V_0]$ as a functional of 
$G$ and  the bare interaction $V_0$ (some examples of this 
procedure are given in the next section). With the
functional $\Gamma [G, V_0 ]$ defined in this way we define
the following $\Phi$ functional
\be
\Phi [G, V_0] = - \Xi [ G, \Gamma [G, V_0] ]
- \frac{i}{4} \tr \left\{ V_0 G_2 [ G, \Gamma [G, V_0]] \right\}
\ee
and the action functional
\be
i Y[ G, V_0 ] = 
-\tr \left\{ \ln ( \Sigma  - G_0^{-1} ) \right\} 
-\tr \left\{ \Sigma G \right\}  + \Phi [ G, V_0 ]
\label{eq:LW}
\ee
where in this expression self-energy $\Sigma [ G, \Gamma [G, V_0]]$ must also
be regarded as a functional of $G$ and $V_0$. 
From the definition of $\Phi$ it then follows directly that
\bea
\delta \Phi &=& -\tr \left\{ \tilde{\Sigma}_C \delta G \right\} + 
\frac{i}{4} \tr \left\{ \tilde V_0 \delta \Gamma \right\} \nonumber \\
&+& \tr \left\{( \Sigma + \tilde{\Sigma}_C) \delta G \right\}
- \frac{i}{4} \tr \left\{ \tilde V_0 \delta \Gamma \right\} \nonumber \\
&=& \tr \left\{ \Sigma \delta G \right\}
\eea
We therefore obtain the result 
\be
\frac{\delta \Phi}{\delta G(12)} = \Sigma (21)
\ee
We further have that the functional $Y[G, V_0]$ of Eq.(\ref{eq:LW})
is stationary when
\bea
0 &=& -\tr \left\{ ( (\Sigma - G_0^{-1})^{-1} + G) \delta \Sigma \right\} \nonumber \\
&& - \tr \left\{ ( \Sigma - \frac{\delta \Phi}{\delta G})\delta G \right\} 
\eea
i.e. whenever the Dyson equation is obeyed for a $\Phi$-derivable
self-energy. On the basis of the work of Baym~\cite{Baym:PR62} we can therefore
conclude that $\Xi$-derivable theories are conserving.

\section{Approximations using the $\Xi$-functional}

\subsection{Practical use of the variational property}

After having discussed the general properties of the functional $Y[ G, \Gamma ]$
we will discuss its use in the calculation of total energies. 
For a given approximation to
$\Xi[G,\Gamma]$ the stationary point of the functional $Y$ 
corresponds to an approximation for the self-energy and the four-point vertex
obtained from a solution of the Dyson equation and of an equation of Bethe-Salpeter type, 
both of which need to be solved to self-consistency. The solution
of these equations for general electronic systems is computationally 
very expensive or impossible. However, if we are only interested
in total energies, then we can use the variational property
of $Y$ and save greatly in computational cost as the full self-consistency
step can then be skipped. 
To illustrate this we let $G$ and $\Gamma$ be self-consistent solutions to
the variational equations and we let $\tilde{G}$ and $\tilde{\Gamma}$ be
approximations to $G$ and $\Gamma$. Then we have that
\bea
Y[ \tilde{G}, \tilde{\Gamma} ] &=& Y[ G, \Gamma ] + \frac{1}{2}
\tr \left\{ \frac{\delta^2 Y}{\delta G \delta G } \Delta G \Delta G \right\}  \nonumber \\
&+&  \tr \left\{ \frac{\delta^2 Y}{\delta G \delta \Gamma } \Delta G \Delta \Gamma \right\} 
\nonumber \\
&+& \frac{1}{2} \tr \left\{ \frac{\delta^2 Y}{\delta \Gamma \delta \Gamma } 
 \Delta \Gamma \Delta \Gamma \right\} + \ldots
\eea
where $\Delta G=\tilde{G}-G$ and $\Delta \Gamma = \tilde{\Gamma}- \Gamma$
are the deviations from the Green function and four-point vertex to the self-consistent
ones. We see that the error we make in $Y$ is only of second order in
$\Delta G$ and $\Delta \Gamma$. We may therefore obtain
rather accurate energies from rather crude inputs.
These expectations were indeed borne out by our earlier
calculations within the $\Phi$-formalism on atoms and
molecules~\cite{Dahlenetal:PRA06}.
Obviously the actual error we make
also depends on how large the second derivatives of functional $Y$ are.
For this reason the Klein and Luttinger-Ward forms of the functional
perform differently. In fact, experience within the
$\Phi$-functional formalism has shown that the Luttinger-Ward 
is more stable than the Klein functional with respect to changes
of the input Green function~\cite{Dahlenetal:PRA06}.

\subsection{Approximate $\Xi$-functionals}

In the following we study some approximate schemes using the $\Xi$-functional
in order to illustrate the formalism discussed in the preceeding sections.
We restrict ourselves here to the two most simplest examples, the self-consistent second order
approximation and the self-consistent $T$-matrix approximation. 
A more advanced approximation, also involving the particle-hole diagrams,
 is discussed in the section on the practical evaluation of the $\Xi$-functional.

The very simplest nontrivial approximation to the we can make to
the $\Xi$-functional is to take $L_{pp}=L_{ph}=L'=0$.
This yields the functional
\bea
\lefteqn{ iY_{2} [ G, \Gamma ] = -\tr \left\{ \ln ( \Sigma - G_0^{-1} ) \right\} 
-\tr \left\{ \Sigma G \right\} }\nonumber \\
 && + \frac{1}{8} \tr \left\{ GG \Gamma GG \Gamma \right \} 
 - \frac{i}{4} \tr \left\{ V_0 G_2 [ G, \Gamma] \right\}
\eea
which we will denote by $Y_2$ since it only involves second order diagrams.
The variational equations yield
\bea
G^{-1} &=& G_0^{-1} - \Sigma [G,\Gamma] \\
0  &=& \frac{1}{4} \tilde{\Gamma} -\frac{i}{4} \tilde{V}_0
\label{eq:var2ndorder}
\eea
which simply implies that $\Gamma= iV_0$ and that
\bea
\lefteqn{ \Sigma [G,V_0 ] (11') 
 = -i \int d(23) V_0(1231') G(32) }\nonumber \\
&+&  \frac{1}{2} \int d(234567)
V_0 (1234) G(36) G(45) \nonumber \\
&& \times V_0 (5671') G(72) 
\eea
This amounts to a self-consistent solution of the Dyson equation with only
second order diagrams.
A fully self-consistent solution of these equations for 
molecules was recently carried out by us~\cite{DahlenvanLeeuwen:JCP05}.
One of the next simplest approximations is obtained by taking 
$L'=L_{ph}=0$ which yields the functional
\bea
iY_{pp} [ G, \Gamma ] &=&  -\tr \left\{ \ln ( \Sigma - G_0^{-1} ) \right\} 
-\tr \left\{ \Sigma G \right\}  \nonumber \\
&& - L_{pp} [ G, \Gamma] + \frac{1}{8} \tr \left\{ GG \Gamma GG \Gamma \right \} \nonumber \\
&&  - \frac{i}{4} \tr \left\{ V_0 G_2 [ G, \Gamma] \right\}
\eea
The variational equations correspond to
\bea
G^{-1} &=& G_0^{-1} - \Sigma [G,\Gamma] \\
0  &=&  -\frac{\delta L_{pp}}{\delta \Gamma} + \frac{1}{4} \tilde{\Gamma} -\frac{i}{4} \tilde{V}_0
\label{eq:varpp}
\eea
where $\Sigma$ is calculated from Eqs.(\ref{eq:self1}) and Eq.(\ref{eq:self2}).
The second variational Eq.(\ref{eq:varpp}) corresponds to
\be
iV_0  =  \Gamma (1+ \frac{1}{2} GG \Gamma )^{-1} .
\ee
This equation can be inverted to give
\be
\Gamma =  (iV_0) (1-  \frac{1}{2} GG (iV_0) )^{-1}
\label{eq:GammaT}
\ee
and expresses the renormalized four-point vertex as a sum of particle-particle 
(direct and exchange) ladder diagrams
in terms of the bare potential $V_0$. 
The corresponding self-energy is then readily obtained from
Eqs.(\ref{eq:self1}) and (\ref{eq:self2}) by inserting 
the $\Gamma$ of Eq.(\ref{eq:GammaT}) in Eq.(\ref{eq:self2}). 
This approximation
is equivalent to the self-consistent $T$-matrix approximation.
It is clear that the set of approximations can be made 
more and more advanced by using more sophisticated
approximations for the $\Xi$-functional.
In the following sections we will discuss the numerical
evaluation of $iY$. We will then among other things, 
consider an approximate four-point vertex
obtained from the $T$-matrix approximation as an approximate input
for the evaluation of the energy functional $iY$ at a more sophisticated
level of perturbation theory.

\section{Practical evaluation of the functional}

\subsection{Evaluation of the traces}

In this section we discuss the how to evaluate the
functional $Y[ G, \Gamma ]$ in actual applications.
Our goal is to evaluate $Y$ for an equilibrium system
in which case all two-time quantities depend on 
relative time variables on the vertical stretch
of the Keldysh contour. In that case it is convenient to
go over to a Matsubara representation
(we use the notation of Kadanoff and Baym~\cite{KadanoffBaym:book})
\bea
A(t-t') &=& \frac{i}{\beta} \sum_z e^{-i z (t-t')} A(z) \\
A(z) &=& \int_0^{-i\beta} dt A(t-t') e^{i z (t-t')}
\eea
where the times are imaginary ( $t=-i\tau$ for $0\leq \tau \leq \beta$)
and where $z= i n\pi/\beta$ are the Matsubara frequencies
which run over even or odd integers $n$ depending on whether $A$ is
a bosonic or fermionic function.
In this way the equation of motion for the Green function
simply attains the form
\bea
\lefteqn{ (z- h(\bx_1)) G(\bx_1 \bx_2,z) = \delta (\bx_1, \bx_2) } \nonumber \\
&+& \int d\bx_3 \Sigma (\bx_1 \bx_3, z) G(\bx_3 \bx_2, z)
\eea
For the traces of two-point functions we have the expression
\bea
\tr A &=& \int_0^{-i\beta}  dt d\bx \, A(1,1^+) \nonumber \\
&=& \lim_{\eta \rightarrow 0^+} \sum_z \int d\bx 
\, e^{\eta z} A(\bx , \bx, z) 
\eea
For the various traces in the functional $Y$ it is further convenient
to introduce a one-particle basis, such that we can write
\be
A(\bx_1, \bx_2, z) = \sum_{ij} A_{ij} (z) \varphi_i (\bx_1) \varphi_j^* (\bx_2)
\ee
Then we have, for instance, that 
\be
\tr AB = \lim_{\eta \rightarrow 0^+} 
\sum_{ij, z}  e^{\eta z} A_{ij}(z) B_{ji} (z)
\ee
If we choose the orbitals to be eigenfunctions of the
one-particle Hamiltonian $h$,
\be
h(\bx) \varphi_i (\bx) = e_i \varphi_i (\bx) 
\ee
then the equation of motion of the Green function attains the form
\be
(z-e_i) G_{ij} (z) = \delta_{ij} + \sum_{k} \Sigma_{ik} (z) G_{kj} (z)
\ee
and we see immediately that the noninteracting Green function $G_0$ is
given by 
\be
G_{0,ij} (z)= \frac{\delta_{ij}}{z-e_i}
\ee
Consequently the grand potential for the
noninteracting system is given by $\Omega_0 = iY_0/\beta$ 
where~\cite{LuttingerWard:PR60,BlaizotRipka:book,Orlewicz:APP87}
\bea
\Omega_0 &=& -\frac{1}{\beta} \tr \ln \left\{ - G_0^{-1} \right\} \nonumber \\
 &=& -\frac{1}{\beta} \lim_{\eta \rightarrow 0^+} \sum_i \sum_z e^{\eta z}
\ln (e_i-z) \nonumber \\
&=& - \frac{1}{\beta} \sum_i \ln (1 + e^{-\beta e_i} )
\eea
In the zero-temperature limit $\beta \rightarrow \infty$ this simply gives
\be
\lim_{\beta \rightarrow \infty} \Omega_0 = \sum_{i=1}^N e_i
\ee
where the sum runs over the $N$ occupied electron orbitals.
Note that the chemical potential $\mu$ is included in
$h$ such that $e_i =\epsilon_i-\mu$ where $\epsilon_i$ are the
eigenvalues of the one-body part of the Hamiltonian.

As a next step we will discuss how to evaluate the functional
on an approximate Green function $\tilde{G}$ and an
approximate vertex $\tilde{\Gamma}$. The input Green
function will in practice not be a fully interacting
Green function but rather one obtained from a
local density approximation (LDA) or
from a Hartree-Fock approximation.
With approximate inputs $\tilde{G}$ and $\tilde{\Gamma}$ 
the first term in Eq.(\ref{eq:LWform})
can be written in a computationally convenient form 
as~\cite{DahlenvonBarth:PRB04}
\bea
\lefteqn{ -\tr \ln \left\{ \Sigma [\tilde{G}, \tilde{\Gamma}]  - G_0^{-1} \right\}
} \nonumber \\
&& = -\tr \left\{ \ln (- \bar{G}^{-1} ) \right\}  - \tr \left\{  \ln ( 1 - \bar{G} \Sigma_C [\tilde{G}, \tilde{\Gamma}] )\right\}
\label{eq:lnconv}
\eea
where we defined 
\be
\Sigma_C [\tilde{G},\tilde{\Gamma} ] = \Sigma [ \tilde{G} , \tilde{\Gamma}] -
\Sigma^{HF}[ \tilde{G} ]
\ee
and the Green function $\bar{G}$
from the Dyson equation
\be
\bar{G} = G_0 + G_0 \Sigma^{HF} [\tilde{G} ] \bar{G}
\ee
The Green function $\bar{G}$ therefore presents the first
iteration towards the Hartree-Fock Green function starting from
$\tilde{G}$. Therefore $\bar{G}= G^{HF}$ when we 
take $\tilde{G}=G^{HF}$ as an input Green function.
The term $\Sigma_C$ represents the correlation part of the
self-energy evaluated at an approximate $\tilde{G}$ and
$\tilde{\Gamma}$.
The reason for introducing $\bar{G}$ is that by doing this we
have in the last term of Eq.(\ref{eq:lnconv})
eliminated a static part of the self-energy, which makes
this term well defined without a 
convergence factor and also makes it decay much faster for large frequencies
which is computationally advantageous as was shown in Ref.~\cite{DahlenvonBarth:PRB04}.
The first term in Eq.(\ref{eq:lnconv})
can be evaluated analytically to
give
\be
i \bar{Y}_0 = -\tr \left\{ \ln (- \bar{G}^{-1} ) \right\}
= -  \sum_i \ln (1 + e^{-\beta \bar{e}_i} )
\ee
where $\bar{e}_i = \bar{\epsilon}_i-\mu$ and $\bar{\epsilon}_i$
are the eigenvalues the Hartree-Fock equations with
a nonlocal self-energy $\Sigma^{HF} [\tilde{G} ]$.
In practice (for instance for LDA input Green functions)
these eigenvalues are close to the true Hartree-Fock
eigenvalues.
Now the functional $Y[\tilde{G}, \tilde{\Gamma} ]$ can be
written as
\bea
\lefteqn{ iY[\tilde{G},\tilde{\Gamma} ] = 
i \bar{Y}_0 } \nonumber \\
&& -\tr \left\{ \ln ( 1  - \bar{G} \Sigma_C [\tilde{G},\tilde{\Gamma} ] ) \right\} 
-\tr \left\{ \Sigma [\tilde{G}, \tilde{\Gamma}] \tilde{G} \right\} \nonumber \\
&& - L [\tilde{G} , \tilde{\Gamma} ] - L' [\tilde{G}, \tilde{\Gamma} ]  - 
\frac{i}{4} \tr \left\{ V_0 G_2 [ \tilde{G}, \tilde{\Gamma} ] \right\}
\eea
The second term can be evaluated by diagonalization of $\bar{G} \Sigma_C$
since for a matrix $A(z)$ we have
\be
\tr \left\{ \ln (1- A) \right\} = \lim_{\eta \rightarrow 0^+} 
\sum_{z,i} e^{\eta z} \ln (1- \lambda_i (z))
\label{eq:freqsum}
\ee
where $\lambda_i (z)$ are the eigenvalues of $A(z)$.
This completes one part of the evaluation of the 
functional $Y$. 

\subsection{Evaluation of the $L'=0$-functional}

Let us now discuss the evaluation of the $L[G, \Gamma]$
and $L' [G, \Gamma ]$ functionals. The evaluation of even
the lowest order term of the $L'$-functional will be computationally
very difficult in practice. The first term in the expansion 
of $L'$ is the pentagon of Fig.(\ref{fig:penta}) containing
five fourvertices $\Gamma$. 
Since every four-point vertex depends on four space-time coordinates
the pentagon is (apart from the spin summations) formally
an $80$-dimensional integral. 
Fortunately, even the approximation
$L'=0$ represents a very sophisticated many-body approximation.
We will therefore in the following 
concentrate on this case and consider the evaluation of the
functional
\bea
iY[G,\Gamma] &=& i\bar{Y}_0 -\tr \left\{ \ln ( 1  - \bar{G} \Sigma_C [G,\Gamma] ) \right\} 
-\tr \left\{ \Sigma G \right\} \nonumber \\
&& - L [G,\Gamma]   - 
\frac{i}{4} \tr \left\{ V_0 G_2 [ G, \Gamma] \right\}
\label{eq:Snew}
\eea
The evaluation of the first terms in this
expression has been discussed in the preceding subsection and 
we will therefore concentrate on evaluation of $L[G,\Gamma]$.
In this term the trace is taken over two-particle functions
and its evaluation will therefore be slightly different from the
case discussed above. 

As our approximate $\Gamma$ we will take 
the sum of all particle-particle and exchange ladders in terms of $V_0$
for which we will eventually take the zero-frequency limit.
This is the approximate $T$-matrix used in Ref.~\cite{KatsnelsonLichtenstein:EPJB02}.
This approximate $\Gamma$ we will denote as $\tilde{\Gamma}$.
This approximate $\Gamma$ will be expressed in terms of 
our approximate Green function which we will denote
with $\tilde{G}$.
Then from Eq.(\ref{eq:GammaJ}) we have
\be
\tilde{\Gamma} = iV_0 + \frac{i}{2} V_0 \tilde{G} \tilde{G} \tilde{\Gamma}
\label{eq:tildegamma}
\ee
If we write
\be
V_0 (1234) = \delta(t_1,t_2) \delta (t_1,t_4) \delta (t_2,t_3)
V_0 (\bx_1 \bx_2 \bx_3 \bx_4)
\ee
with $V_0 (\bx_1 \bx_2 \bx_3 \bx_4)$ explicitly given in
Eq.(\ref{eq:V0def}),
we see that we can write
\be
\tilde{\Gamma}(1234) = \delta (t_1,t_2) \delta (t_3,t_4) \gamma
(\bx_1 \bx_2 \bx_3  \bx_4; t_1 t_3)
\ee
If we further expand $\gamma$ in a basis
as
\bea
\lefteqn{ \gamma (\bx_1 \bx_2 \bx_3  \bx_4; t_1 t_3)
= \sum_{ijkl} \gamma_{ijkl}(t_1 t_3) } \nonumber \\
&\times& \varphi_i^* (\bx_1) \varphi_j^* (\bx_2)
\varphi_k (\bx_3) \varphi_l (\bx_4)
\eea
then from Eq.(\ref{eq:tildegamma}) we see that $\gamma$
satisfies
\bea
\lefteqn{ \gamma_{ijkl} (t_1 t_3) = i \delta (t_1,t_3) 
V_{0,ijkl} }\nonumber \\
&&  + \frac{i}{2} \sum_{pqrs} 
\int_{0}^{-i\beta} dt_2  V_{0,ijpq} \nonumber \\
&& \times \tilde{G}_{qr} (t_1, t_2) \tilde{G}_{ps} (t_1, t_2)
\gamma_{rskl} (t_2 t_3)
\eea
which in frequency space attains the form
\bea
\lefteqn{ \gamma_{ijkl} (z) = i  
V_{0,ijkl} 
 - \frac{1}{2 \beta} \sum_{z_1} \sum_{pqrs} 
 V_{0,ijpq} } \nonumber \\
&& \times \tilde{G}_{qr} (z_1) \tilde{G}_{ps} (z-z_1)
\gamma_{rskl} (z)
\eea
(note that for $\gamma$ we have to sum over the
even Matsubara frequencies). For simple approximate
Green functions $\tilde{G}$ of Hartree-Fock or
local density type the frequency sum over $z_1$ is readily evaluated.
We are now ready to evaluate the functionals $L_{pp}[ \tilde{G} , \tilde{\Gamma}]$ 
and $L_{ph} [ \tilde{G}, \tilde{\Gamma} ]$. They are
given by the expressions
\bea
L_{pp} &=& \tr \left\{ \ln (1+ A) \right\} -\tr \left\{ A \right\} + \frac{1}{2} 
\tr \left\{ A^2 \right\} \\
L_{ph} &=& \frac{1}{2}\tr \left\{ \ln (1- B) \right\} +
\frac{1}{2} \tr \left\{ B \right\} + \frac{1}{4} 
\tr \left\{ B^2 \right\} 
\eea 
where $A=GG\Gamma$ and $B= \widehat{GG} \bar{\Gamma}$.
Therefore in order to calculate $L_{pp}$
and $L_{ph}$ we have to diagonalize $A$
and $B$ in a two-particle basis. Let us
start by the calculation of $A$.
We have for our approximate $\tilde{\Gamma}$ and
$\tilde{G}$ : 
\bea
A_{ijkl} (t_1 t_2 t_3 t_4) &=& \delta (t_3,t_4)  \sum_{pq}
\int_0^{-i\beta} dt_5 \tilde{G}_{ip}(t_1,t_5) \nonumber \\
&& \times 
\tilde{G}_{jq} (t_2,t_5) \gamma_{pqkl} (t_5 t_3)
\label{eq:Amatrix}
\eea
Because of the equal-time delta function in Eq.(\ref{eq:Amatrix})
we find that
\bea
\lefteqn{ \tr \left\{ A^n \right\} = } \nonumber \\
&& = \int d(11'\ldots nn') \langle 11' | A| 22' \rangle \ldots \langle nn'| A | 11' \rangle
\nonumber \\
&& = \sum_{p_1 \ldots p_n} \int_0^{-i\beta}
d(t_1 \ldots t_n) \bar{A}_{p_1 p_2} (t_1,t_2) \ldots \bar{A}_{p_n p_1} (t_n,t_1) \nonumber \\
&& = \lim_{\eta \rightarrow 0^+} \sum_z e^{\eta z}
 \bar{A}_{p_1 p_2} (z) \ldots \bar{A}_{p_n p_1}(z)
\eea
where $p_k= (i_k j_k)$ are multi-indices and
where we defined
\bea
\bar{A}_{ijkl} (t_1 t_3) &=&  \sum_{pq}
\int_0^{-i\beta} dt_5 \tilde{G}_{ip}(t_1,t_5) \nonumber \\
&& \times 
\tilde{G}_{jq} (t_1,t_5) \gamma_{pqkl} (t_5 t_3)
\label{eq:barAmatrix}
\eea
which in frequency space attains the form
\bea
\bar{A}_{ijkl} (z) = \frac{i}{\beta}
\sum_{z_1, pq} \tilde{G}_{ip}(z_1) \tilde{G}_{jq} (z-z_1)
\gamma_{pqkl} (z)
\label{eq:barAfreq}
\eea
From diagonalization of $\bar{A}_{p q} (z)$ where $p=(ij)$ and $q=(kl)$ we
then immediately obtain
\bea
\lefteqn{ L_{pp} [ \tilde{G}, \tilde{\Gamma} ] = } \nonumber \\
&& \sum_{z,p}  ( \ln (1+ \lambda_p (z))  -\lambda_p (z) + \frac{1}{2} 
\lambda_p^2 (z) )
\eea
where $\lambda_p (z)$ are the eigenvalues of $\bar{A}(z)$.
Let us finally concentrate on the evaluation of $B$.
This expression is given by
\bea
B_{ijkl} (t_1 t_2 t_3 t_4) &=& \sum_{pq}
\tilde{G}_{qi} (t_4,t_1) \nonumber \\
&& \times \tilde{G}_{jp} (t_2,t_3)
\gamma_{pklq} (t_3 t_4)
\eea
This expression depends on three relative times which makes
it awkward to evaluate the logarithm. We
therefore follow reference \cite{KatsnelsonLichtenstein:EPJB02} and replace 
in frequency space $\gamma_{ijkl}(z)$ by its zero-frequency 
limit $\gamma_{ijkl}(0)$,
\be
\gamma_{ijkl}(t_3 t_4) = \gamma_{ijkl} (0) \delta (t_3,t_4)
\ee
such that
\bea
B_{ijkl} (t_1 t_2 t_3 t_4) &=& \delta (t_3 t_4) \sum_{pq}
\tilde{G}_{qi} (t_4,t_1) \nonumber \\
&& \times \tilde{G}_{jp} (t_2,t_3)
\gamma_{pklq} (0)
\eea
Then, similarly as for the quantity $A$ we have
\bea
 \tr \left\{ B^n \right\} = 
\lim_{\eta \rightarrow 0^+} \sum_z e^{\eta z}
 \bar{B}_{p_1 p_2} (z) \ldots \bar{B}_{p_n p_1}(z)
\eea
where
\be
\bar{B}_{ijkl} (z) = \frac{i}{\beta} 
\sum_{z_1, pq}  \tilde{G}_{qi} (z_1) \tilde{G}_{jp} (z_1+z) \gamma_{pklq}(0)
\label{eq:barBmatrix}
\ee
Now $\bar{B}(z)$ is readily diagonalized with respect to its
two-particle indices to give
\bea
\lefteqn{ L_{ph} [ \tilde{G}, \tilde{\Gamma} ] = } \nonumber \\
&& \sum_{z,p}  ( \frac{1}{2} \ln (1 - \hat{\lambda}_p (z))  + \frac{1}{2} 
\hat{\lambda}_p (z) + \frac{1}{4} 
\hat{\lambda}_p^2 (z) )
\eea
where $\hat{\lambda}_p (z)$ are the eigenvalues of $\bar{B} (z)$.
The full functional $L[ \tilde{G}, \tilde{\Gamma} ]$ is
then constructed as 
\be
L [ \tilde{G}, \tilde{\Gamma} ] = L_{pp} [ \tilde{G}, \tilde{\Gamma} ] 
+ L_{ph} [ \tilde{G}, \tilde{\Gamma} ] -\frac{1}{8} \tr \left\{ A^2 \right\}
\ee
where the last term is easily found by summing the squares of the
eigenvalues of $A$ and performing a frequency sum.
It finally remains to calculate an explicit expression for
$\Sigma [\tilde{G}, \tilde{\Gamma} ]$ and to evaluate the
last term in Eq.(\ref{eq:Snew}).
The self-energy is readily calculated 
from Eqs.(\ref{eq:self1}) and (\ref{eq:self2}) to be
\be
\Sigma_{ij}(z) = \Sigma_{ij}^{HF} + \Sigma_{ij,C} (z)
\ee
where
\be
\Sigma_{ij}^{HF} = \frac{1}{\beta} \lim_{\eta \rightarrow 0^+}
\sum_{z} V_{0,ipqj} \tilde{G}_{qp}(z)
\ee
and
\bea
\Sigma_{ij,C}(z) &=& \frac{i}{2\beta} 
\sum_{z_1,z_2} \sum_{pqrstu} V_{0,ipqr}
\tilde{G}_{rs} (z_1) \tilde{G}_{qt} (z_2) \nonumber \\
&& \times 
\tilde{G}_{up}(z_1+z_2-z) \gamma_{stuj}(0)
\eea
This expression is, of course, considerably simplified 
when we use a diagonal input Green function. 
This finally concludes the discussion on the
practical evaluation of the functional.

In summary: evaluation of the functional $Y$ in practice
therefore essentially involves the
diagonalization of the one-particle matrix $A(z)$ of
Eq.(\ref{eq:freqsum}) and the diagonalization 
of the matrices $\bar{A} (z)$ and $\bar{B}(z)$ of 
Eqns.(\ref{eq:barAmatrix})
and (\ref{eq:barBmatrix}) in a two-particle basis
followed by a frequency summation.
This is, for instance within the DMFT approach used 
by Katsnelson and Lichtenstein~\cite{KatsnelsonLichtenstein:EPJB02},
a numerically quite feasible procedure.

\section{Conclusions}

In this work we studied variational functionals of the
Green function and the renormalized four-point vertex in order to
calculate total energies for strongly correlated systems.
The variational functionals were derived by Legendre
transform techniques starting from an expression
of the action (or grand potential) defined on the
Keldysh contour. The structure of the functionals was
further analyzed by means of diagrammatic techniques.
We finally gave a detailed discussion of the practical use
and evaluation of these for different approximate 
functionals. Future applications along the lines 
described are intended.

Finally we comment on further applications
of the variatonal functionals.
It was found that within the $\Phi$ and the $\Psi$-formalism
could be succesfully used to derive expressions for
response functions within
time-dependent density-functional theory (TDDFT)~\cite{Ulfetal:PRB05}.
This was done by inserting approximate 
Green  functions $G[v]$, coming from
a noninteracting system with a local potential $v$, 
into the variational functionals. Then the
potentials were optimized by requiring that 
$\delta Y/\delta v=0$. Due to the one-to-one
correspondence between the density and the potential
(as follows from 
the time-dependent generalization of the Hohenberg-Kohn theorem~\cite{DreizlerGross:book})
this then implies that we are optimizing a time-dependent
density functional. The optimized potentials are then to be interpreted
as Kohn-Sham potentials. In this way one obtains a density functional
for every diagrammatic expression from the $\Phi$- or $\Psi$-functional.
A similar procedure can now be carried out for the $\Xi$-functional.

A further point of future investigation is concerned with
finding the variationally most stable functional. It was already
mentioned that the Klein and Luttinger-Ward (LW) forms of the
functional lead to different results. The
Luttinger-Ward form was found to be more stable.
This is probably due to the fact that the second derivatives
of the LW functional are smaller than those of the Klein functional.
However, it is very well possible that one could derive a better
functional that would make the second derivatives even smaller
or make them vanish. In that case the errors we make would be
only to third order in the deviation $\Delta G$ of the 
input Green to the true self-consistent one.
This still remains an issue for future investigations.
Finally we mention that work on implementation of the formalism discussed
here is in progress.

\section{Acknowledgments}
We like to thank Prof.M.I.Katsnelson and Prof.A.I.Lichtenstein
for useful discussions and for interest in this work.

\appendix

\section{A generating functional for the Green function}
\label{genfunc}

In order to obtain the Green functions as variational derivatives
we define~\cite{Baym:PR62} an evolution operator in terms of
a time- and space nonlocal potential $u(12)$
\bea
\lefteqn{ \hU [u] (t_0-i\beta, t_0) = 
T_C \exp (-i \int dt\hH (t) - } \nonumber \\
&& i   \int d1 \int d2 \, \psic (\bx_1) u(12) \psia  (\bx_2) )
\eea
where we used the compact notation $1=(\bx_1, t_1)$.
Since we have now two times in the exponent this expression only
has meaning if we define how the time-ordering is specified if we expand
this expression. It is defined as follows:
\bea
\lefteqn{ \hU [u] (t_0-i\beta,t_0) \equiv \hU [u=0] (t_0-i\beta,t_0)  + } \nonumber \\
&& \sum_{n=1}^{\infty} \frac{(-i)^n}{n!} \int d(1 1' \ldots n n') \,
u(1'1) \ldots u(n'n) \nonumber \\
&& \times \langle T_C[ \psic_H (1') \psia_H (1) \ldots 
\psic_H (n') \psia_H (n) ] \rangle 
\label{eq:Uexpand}
\eea
where the expectation values under the integral sign are averages
(as in Eq.(\ref{eq:average})) in the absence of the nonlocal field $u$.
This definition agrees in the limit of a time-local potential, i.e.
$u(12)=u(\bx_1 t_1, \bx_2 t_1)\delta (t_1^+,t_2)$, with an
expression that can be derived directly from the time-dependent
Schr\"odinger equation.
We now define the functionals
\bea
Z[u] &=& \Tr \left\{ U[u](t_0 -i\beta, t_0) \right\} \\
iY[u] &=& - \ln Z[u]
\eea
Then the one-particle Green function in the presence of
the nonlocal field $u$ is defined as:
\bea
G_u (11') \equiv i \frac{\delta Y}{\delta u(1'1)} = 
- \frac{1}{Z[u]} \frac{\delta Z}{\delta u(1'1)} 
\label{eq:Gudef}
\eea
When evaluated at $u=0$ the Green function reduces to the familiar one
\be
G_{u=0}(11') = -i \langle T_C [ \psic_H (1) \psia (1')]\rangle 
\ee 
Let us note that one should be careful with dealing with
time-nonlocal potentials.
It would, for instance, be tempting to think that
$G_u$ would be given by the expression
\bea
\lefteqn{G_u (11') = } \nonumber \\
&& -i \frac{\Tr \left\{ U[u](t_0-i\beta, t_0) T_C [ \psic_H (1') \psia_H (1) ] \right\}}
{\Tr \left\{ U[u](t_0 -i\beta, t_0) \right\}} 
\label{eq:Gudef2}
\eea
where the Heisenberg operators in the presence of $u$ would be given by 
$\hat{O}_H = U[u](t_0,t) \hat{O} \hU [u] (t,t_0)$. However, this expression
is {\em not} valid if $u$ is nonlocal in time. For instance, when expanding
Eq.(\ref{eq:Gudef}) and (\ref{eq:Gudef2}) in powers of $u$ 
using Eq.(\ref{eq:Uexpand}) one immediately sees
that certain time-orderings of the field operators in Eq.(\ref{eq:Gudef})
are absent in Eq.(\ref{eq:Gudef2}). Similarly the evolution operator
$\hat{U}[u](t,t')$ does not satisfy a simple equation of motion.
However, we can still derive the equations of motion for $G_u$ 
on the basis of the hierarchy equations of the $n$-body Green functions
in the absence of the nonlocal field $u$, as we
will show below. \\
More generally we can now define $n$-body Green functions $G_{n,u}$ 
from a repeated
differentiation of $Z[u]$, i.e.
\bea
\lefteqn{ \frac{1}{Z[u]} \frac{\delta^n Z}{\delta u(1' 1)\ldots \delta u(n' n) } } \nonumber \\
&=& \epsilon_n G_{n,u} (1 \ldots n, 1' \ldots n')
\label{Gundef}
\eea
where $\epsilon_n = (-1)^{n(n+1)/2}$. The prefactor $\epsilon_n$
results from reordering the operator product
$\langle T_C[ \psic (1') \psia (1) \ldots \psic (n') \psia (n)]\rangle$
to $\langle T_C [ \psia (1) \ldots \psia (n) \psic (1')
\ldots \psic (n') ] \rangle$ as is easily verified by induction. One can readily check that for $u=0$
Eq.(\ref{Gundef}) agrees with our previous definition of the
$n$-body Green function of Eq.(\ref{eq:gndef}). 
From Eq.(\ref{Gundef}) we further immediately obtain that
\bea
\lefteqn{ \frac{\delta G_u (14)}{\delta u(32)} = \frac{\delta}{\delta u(32)}
\big( - \frac{1}{Z[u]} \frac{\delta Z}{\delta u(41)} \big) } \nonumber \\
 && = \frac{1}{Z^2} \frac{\delta Z}{\delta u(14)} \frac{\delta Z}{\delta u(32)}
- \frac{1}{Z} \frac{\delta^2 Z}{\delta u(41) \delta u(32)} \nonumber \\
&& = G_u (14) G_u (23) - G_{2,u} (1234) 
\eea
This is Eq.(\ref{eq:Gderiv}) used in section \ref{sec:hedin}.
As a next step we derive the hierarchy equations for the Green functions $G_{n,u}$.
From Eq.(\ref{Gundef}) we see that we can expand the functional $Z[u]$ as a Taylor series expansion in $u$ as
\bea
\lefteqn{ Z[u]= Z[0] \sum_{n=0}^{\infty} \frac{\epsilon_n}{n!} \int d(1 1' \dots n n') }
\nonumber \\
&& \times G_n(1 \dots n,1' \dots n')
u(1'1) \dots u(n' n)
\label{eq:Zexp}
\eea
where the term with $n=0$ is just defined to be one. The Green functions $G_n$ are the Green functions
in the absence of the field $u$. The one-body and $n$-body Green functions can therefore be expressed in terms
on the field-free Green functions using Eqs.(\ref{eq:Gudef}), (\ref{Gundef}) and (\ref{eq:Zexp}). One obtains
for $G_u$ and $G_{2,u}$ the equations
\bea
\lefteqn{ \frac{Z[u]}{Z[0]} G_u (11') =  G(11') }\nonumber \\
&-& \sum_{n=2}^{\infty} \frac{\epsilon_n}{(n-1)!} 
\int d(2 2' \ldots n n') \nonumber \\
&& \times G_n (1\ldots n, 1' \ldots n') u(2'2) \ldots u(n'n) 
\label{eq:GuG}\\
\lefteqn{ \frac{Z[u]}{Z[0]} G_{2,u} (12;1'2') =  G(12;1'2') } \nonumber \\
&-& \sum_{n=3}^{\infty} \frac{\epsilon_n}{(n-2)!}
\int d(3 3' \ldots n n') \nonumber \\
&& \times G_n (1\ldots n, 1' \ldots n') u(3'3) \ldots u(n'n)
\label{eq:G2uG}
\eea
From these expressions we see that $G_u$ and $G_{2,u}$ inherit the
Kubo-Martin-Schwinger boundary conditions from the $G_n$.
If we act with the operator $i\prt_{t_1}-h(1)$ on both sides of Eq.(\ref{eq:GuG}) and
use the Martin-Schwinger hierarchy equations Eq.(\ref{eq:hierarchy})
for the Green functions $G_n$ in the absence of the $u$-field 
together with Eq.(\ref{eq:G2uG}) we 
obtain, after slightly tedious but straightforward
manipulations, the equation of motion for $G_u$
\bea
\lefteqn{ (i\prt_{t_1}-h (1))G_u (11') = \delta (11') } \nonumber \\
 && + \int d2 u(12) G_u (21') \nonumber \\
&&-  i \int d\bx v(\bx_1 ,\bx) G_{2,u} (1,\bx t_1, \bx t_1^+ ,1')
\label{eq:motionGu}
\eea
By functional differentiation with respect to $u$ we can generate
equations of motion for the higher-order Green functions.
To see this we first multiply (\ref{eq:motionGu}) by $Z[u]$
to obtain
\bea
\lefteqn{ (i\prt_{t_1}-h (1))(-i) \langle 1 1' \rangle  = \delta (11') Z[u] }
\nonumber \\ && + 
\int d\bar{1} u(1\bar{1}) (-i) \langle \bar{1} 1' \rangle \nonumber \\
&& - i \int d\bx v(\bx_1 ,\bx) (-i)^2 \langle 1,\bx t_1, \bx t_1^+ ,1' \rangle
\label{eq:motion2}
\eea
where we introduced the simplified notation
\bea
 \langle 1 \ldots n; 1' \ldots n' \rangle =
Z[u] \, G_{n,u}(1\ldots n,1' \ldots n')
\eea
We use the convention that the primed variables are always associated
with creation operators and that the unprimed variables are always
associated with the annihilation operators. Taking the
functional derivative of Eq.(\ref{eq:motion2}) with respect to $u(2'2)$
then gives
\bea
\lefteqn{ (i\prt_{t_1}-h (1))(-i)^2 \langle 1 1' 2' 2 \rangle  = 
\delta (11') (-i) \langle 2' 2 \rangle } \nonumber \\
&& + \delta (12') (-i) \langle 2 1' \rangle 
\nonumber \\
&&+ \int d\bar{1} u(1\bar{1}) (-i)^2 \langle \bar{1} 1' 2' 2 \rangle 
\nonumber \\
&& - i \int d\bx v(\bx_1 ,\bx) (-i)^3 \langle 1,\bx t_1, \bx t_1^+ ,1' 2' 2\rangle
\eea
Reordering the indices and dividing by $Z[ u]$ then gives
\bea
\lefteqn{ (i\prt_{t_1}-h (1)) G_{2,u} (12 1' 2')  = } \nonumber \\
&& -\delta (11') G_u ( 2 2')  + \delta (12') G_u ( 2 1' ) \nonumber \\
&&+ \int d\bar{1} u(1\bar{1}) G_{2,u} ( \bar{1} 2 1' 2' ) \nonumber \\
&& - i \int d\bx v(\bx_1 ,\bx) G_{3,u}( 12,\bx t_1 , \bx t_1^+ ,1' 2' )
\eea
By continued differentiation we obtain the general hierarchy equations
for $G_{n,u}$
\bea
\lefteqn{ (i\prt_{t_1} - h (1)) G_{n,u} (1 \ldots n , 1' \ldots n') } \nonumber \\
&=& \sum_{j=1}^n \delta (1 j')(-1)^{n-j} 
G_{n-1,u} (2 \ldots n , 1' \dots j'-1 , j'+1 \dots n') \nonumber \\
&+& \int d\bar{1} u(1 \bar{1}) G_{n,u} (\bar{1} 2 \ldots n, 1' \ldots n') \nonumber \\
&&
- i \int d\bx v(\bx_1 , \bx) G_{n+1, u} (1 \ldots n, \bx t_1 , \bx t_1^+, 1' \dots n')
\label{eq:hierarchyU}
\eea
These equations are readily checked by induction if we multiply them
by $Z[u]$ and take the functional derivative with respect to $u(n'+1, n+1)$.
We have therefore established that Green functions $G_{n,u}$
satisfy an obvious generalisation of the hierarchy equations.
The relation Eq.(\ref{eq:hierarchyU}) is the main result of this
Appendix and will be essential to the derivation in the next section. 
Note further that Eq.(\ref{eq:hierarchyU}) can be used to derive a Wick's
theorem in the presence of the nonlocal field $u$. If we 
put $w=0$ we find that the noninteracting $n$-body Green
functions $G_{n,u}$ satisfy Eq.(\ref{eq:hierarchyU}) if they are written
as determinants in terms of $G_{u}$.

\section{The equation of motion of $G_{u,V}$}
\label{genfunc2}

The main goal in this Appendix is to
derive the equations of motion 
for the Green function
in the presence of the a nonlocal one-body potential $u(12)$
and a nonlocal two-body potential $V(1234)$.
As in Appendix~\ref{genfunc} the main difficulty is caused
by the fact that $u$ and $V$ are nonlocal in time. 
Our final result can be obtained with help of Eq.(\ref{eq:hierarchyU}).
Let $Z[u,V]$ be given by
\be
Z[u,V] = \Tr \left\{ \hU [u,V](t_0-i\beta, t_0 ) \right\}
\ee
where
\bea
\lefteqn{ \hU [u,V] (t_0-i\beta, t_0) = 
T_C \exp (-i \int dt\hH_0 (t)  }\nonumber  \\ 
&& -i  \int d1 \int d2 \, \psic (\bx_1) u(12) \psia  (\bx_2) \nonumber \\
&&- \frac{i}{4} \int d(1234) V(1234) \nonumber \\
&& \times  \hpsid (\bx_1) \hpsid (\bx_2 ) \hpsi (\bx_3) \hpsi (\bx_4)  )
\eea
Due to the multiple time-integrals this expression has only meaning
when we define how the time-ordering is specified when we expand this
expression. We define
\bea
\lefteqn{ \hat{U} [u,V] (t_0-i\beta,t_0) \equiv \hat{U}[0,0] (t_0-i\beta,t_0) + }
\nonumber \\
&& \sum_{n,m=1}^{\infty} \frac{(-i)^{n+m}}{n!m! 4^m} \int d(y_{\bar{1}} \ldots y_{\bar{n}} )
d(X_1 \ldots X_m)  \nonumber \\ 
&& \times u_{\bar{1}} \ldots u_{\bar{n}} 
 V_1 \ldots X_m  T_C [ \hat{y}_{\bar{1}} \ldots \hat{y}_{\bar{n}}
\hat{X}_1 \ldots \hat{X}_m ] 
\label{eq:UuVdef}
\eea
where for the coordinates we introduced the short notation
\bea
y_{\bar{i}} &=& (\bar{i}', \bar{i}) \\
X_i &=& ( (2i-1)', (2i)', 2i-1, 2i) 
\eea
and we further defined
\bea
u_{\bar{i}} &=& u(y_{i}) \\
\hat{y}_{\bar{i}} &=& \psic_H (\bar{i}') \psia_H (\bar{i}) \\
V_i &=& V(X_i) \\
\hat{X}_i &=& \psic_H ((2i-1)') \psic_H ( (2i)') \psia_H (2i-1) \psia_H (2i)
\eea
where the Heisenberg representation of the operators is defined with respect to $\hH_0$.
We then define the $n$-particle Green function $G_{n,u,V}$ in the presence of
the time-nonlocal fields $u$ and $V$ as
\bea
\lefteqn{ \epsilon_n G_{n,u,V} (1 \ldots n; 1'\ldots n') \equiv}  \nonumber \\
&& \frac{1}{Z[u,V]} \frac{\delta^n Z[u,V]}{\delta u(1'1) \ldots u(n'n)}
\eea
where $\epsilon_{n}=(-1)^{n(n+1)/2}$.
For $V=0$ this definition agrees with the definition (\ref{Gundef})
in Appendix~\ref{genfunc} in the absence of two-particle interactions. 
Now from Eq.(\ref{eq:UuVdef}) one can readily derive that
\bea
\lefteqn{ \frac{\delta^k Z[u,V]}{\delta V_1 \ldots \delta V_k} =}  \nonumber \\
&& \frac{(-i)^k}{4^k} \frac{\delta^{2k} Z[u,V]}
{\delta u(1'1) \ldots u((2k)',2k)}
\eea
With this equation we find that we can express the Green functions
equivalently as
\bea
\lefteqn{ G_{2n,u,V} (1 \ldots 2n; 1' \ldots (2n)') }\nonumber \\
&=& \frac{4^n (-i)^n}{Z[u,V]} \frac{\delta^k Z[u,V]}
{\delta V_1 \ldots \delta V_n} 
\label{eq:uVexp1}\\
\lefteqn{ G_{2n+1,u,V} (\bar{1},1,\ldots 2n; \bar{1}',1' \ldots (2n)') } \nonumber \\
&=&- \frac{4^n (-i)^n}{Z[u,V]} \frac{\delta^{2n+1} Z[u,V]}
{\delta u_{\bar{1}} V_1 \ldots V_n}
\label{eq:uVexp2}
\eea
As a particular case we have
\bea
G_{u,V} (11') &=& - \frac{1}{Z[u,V]} \frac{ \delta Z[u,V]}{\delta u(1'1)} \\
G_{2,u,V} (121'2') &=& - \frac{ 4i}{Z[u,V]} \frac{\delta Z[u,V]}{\delta V(1'2'12)}
\eea
These equations are two basic starting equations
Eq.( \ref{eq:G1derv}) and (\ref{eq:G2derv})  of section \ref{sec:varfunc}.
It remains to show that they are related by an equation of motion.
From Eqs.(\ref{eq:uVexp1}) and (\ref{eq:uVexp2}) we see that
 $Z[u,V]$ has the following Taylor expansion around $V=0$.
\bea
\lefteqn{ \frac{Z[ u, V]}{Z[ u,0 ]} =
\sum_{n=0}^{\infty} \frac{i^n}{n! 4^n} 
\int d(X_1 \ldots X_n) } \nonumber \\
&& \times G_{2n,u} (1 \ldots 2n; 1' \ldots 2n')  V_1 \ldots V_n
\label{eq:ZuVexp}
\eea
where we denote $G_{n,u}=G_{n,u,V=0}$.
Similarly for $G_{u,V}$ we have from Eq.(\ref{eq:uVexp2}) the Taylor series expansion
\bea
\lefteqn{ \frac{Z[ u, V]}{Z[ u,0 ]} G_{u,V} (\bar{1} \bar{1}') 
= G_u (\bar{1}, \bar{1}')} \nonumber \\
&+&
\sum_{n=1}^{\infty} \frac{i^n}{n! 4^n}
\int d(X_1 \ldots X_n) \nonumber \\ 
&& \times G_{2n+1,u} (\bar{1} 1 \ldots 2n;  \bar{1}' 1' \ldots 2n')  V_1 \ldots V_{n}
\label{eq:GUV}
\eea
and for the two-particle Green function from Eq.(\ref{eq:uVexp1})
\bea
\lefteqn{\frac{Z[u, V]}{Z[u,0]} G_{2,u,V}(\bar{1} \bar{2}, \bar{1}' \bar{2}') =
G_{2,u}(\bar{1} \bar{2}, \bar{1}' \bar{2}')} \nonumber \\
 &-& i \sum_{n=2}^{\infty} \frac{i^n}{4^{n-1} (n-1) !}
\int d(X_1 \ldots X_{n-1})    \nonumber \\
&& \times 
G_{2n,u} (\bar{1} \bar{2} 1 2 \ldots 2(n-1);  \bar{1}' \bar{2}' 1' 2'  \ldots 2(n-1)') 
\nonumber \\ && \times V_1 \ldots V_{n-1}
\label{eq:G2UV}
\eea
To obtain an equation of motion of $G_{u,V}$ we can use the
hierarchy equations for $G_{n,u}$ of Eq.(\ref{eq:hierarchyU}) for $w=0$
\bea
\lefteqn{ (i\prt_{t_1} - h (1)) G_{n,u} (1 \ldots n , 1' \ldots n') } \nonumber \\
&=& \sum_{j=0}^{n-1} \delta (1 j')(-1)^{n-j} \nonumber \\
&& \times 
G_{n-1,u} (2 \ldots n , 1' \ldots (j-1)', (j+1)' \ldots n') \nonumber \\
&+& \int d\bar{1} \, u(1 \bar{1}) G_{n,u} (\bar{1} 2 \ldots n, 1' \ldots n') 
\label{eq:hierarchyU3}
\eea
If we act with $i \prt_{t_{\bar{1}}} - h (\bar{1})$ on both sides of
Eq.(\ref{eq:GUV}) and subsequently use Eqs.(\ref{eq:hierarchyU3}),(\ref{eq:ZuVexp})
and (\ref{eq:G2UV}) we obtain
\bea
\lefteqn{ (i \prt_{t_{\bar{1}}} - h(\bar{1}) ) 
G_{u,V}(\bar{1} \bar{1}') = \delta (\bar{1}\bar{1}') }
\nonumber \\ 
&& + \int d2 u(\bar{1}2) G_{u,V} (2\bar{1}') \nonumber \\
&& - \frac{i}{2} \int d(234) V ( \bar{1} 2 3 4) G_{2,u,V} (4 3 2 \bar{1}')
\eea
This is the equation of motion for the Green function used in
section \ref{sec:varfunc}. Again by differentiating this equation with
respect to $u$ we obtain the hierarchy equations for the
higher order Green functions $G_{n,u,V}$.

\section{Feynman rules for the two-particle Green function}
\label{section:frules}

In this section we give a brief summary of the
Feynman rules for the two-particle Green function $G_2$
within the Hugenholtz diagram 
technique~\cite{Hugenholtz:57,Nozieres:book, NegeleOrland:book,BlaizotRipka:book}.
The general structure of the two-particle Green function
is as given in Fig.(\ref{fig:four-point vertex}).
The Green function $G_2 (1234)$ is written with the
points $(1234)$ positioned clockwise on four corners of the diagram
where corners 1 and 2 are connected to outgoing lines and
corners 3 and 4 are connected to ingoing lines.
If one expands the evolution operators in the definition of
$G_2$ in powers of the interaction $V$ one finds
for the diagrams the following rules

\begin{enumerate}
\item Every Green function line (contraction according to
Wick's theorem) gives a factor $iG$ .

\item Every vertex gives a factor $-iV$ .
\item Every closed loop of Green function lines gives a minus sign, i.e. we have a
factor $(-1)^l$ where $l$ is the number of closed loops. To find the number of
loops one must replace the Hugenholtz vertex by the first term on the right hand side
of Fig.\ref{fig:hugvertex} (with the same labeling) 
and count the number of loops that appear in this way.
\item A line starting at $3$ and ending at $1$ gives a 
minus sign, i.e. we have a factor $(-1)^{L_{13}}$ where
$L_{13}=1$ when 1 and 3 are connected and zero otherwise.
To determine the connectivity it is necessary 
that one again first replaces the Hugenholtz vertex
by the first term on the right hand side of Fig.\ref{fig:hugvertex}
(see also~\cite{NegeleOrland:book}).
\item Two Green function lines (so-called equivalent lines)
starting from a given vertex
and ending both on the same vertex give a factor $\frac{1}{2}$, i.e.
we have a factor $2^{-p}$ where $p$ is the number of equivalent lines.
\item  There is a factor $(-i)^2$ from the definition of $G_2$.  
\end{enumerate} 

From these rules we find that the overall prefactor of a $G_2$-diagram with
$n$ vertices is given by $i^n (-1)^{l+ L_{13}} 2^{-p}$.
For example, diagrams $(a)-(f)$ in Fig.\ref{fig:G2} have prefactors
$1,-1,i,-\frac{1}{2}, 1$ and $-1$ respectively.

 \end{document}